\documentclass[reprint,superscriptaddress,aps,prl,floatfix]{revtex4-1}

\usepackage{mathtools}
\usepackage{graphicx}
\usepackage{hyperref}
\usepackage[export]{adjustbox}
\usepackage[dvipsnames]{xcolor}
\usepackage{tikz}
\usepackage{braket}
\usepackage{textcomp,gensymb}
\usepackage{dsfont}
\usepackage{comment}

\newcommand{\bvec}[1]{\boldsymbol #1}
\newcommand{\Ucrit}{U_\mathrm{c}}
\newcommand{\VHr}{\ensuremath{V_{\mathrm{H}}(\mathbf{r})}}
\DeclareMathOperator{\hadamard}{\circ}

\begin{document}

\title{Flat bands, electron interactions and magnetic order in magic-angle mono-trilayer graphene}

\author{Zachary A. H. Goodwin}
\thanks{These authors contributed equally to this work.}
\affiliation{Departments of Materials and Physics and the Thomas Young Centre for Theory and Simulation of Materials, Imperial College London, South Kensington Campus, London SW7 2AZ, UK\\}
\author{Lennart Klebl}
\thanks{These authors contributed equally to this work.}
\affiliation{Institute for Theory of Statistical Physics, RWTH Aachen University, and JARA Fundamentals of Future Information Technology, 52062 Aachen, Germany\\}
\author{Valerio Vitale}
\affiliation{Departments of Materials and Physics and the Thomas Young Centre for Theory and Simulation of Materials, Imperial College London, South Kensington Campus, London SW7 2AZ, UK\\}
\author{Xia Liang}
\affiliation{Departments of Materials and Physics and the Thomas Young Centre for Theory and Simulation of Materials, Imperial College London, South Kensington Campus, London SW7 2AZ, UK\\}
\author{Vivek Gogtay}
\affiliation{Departments of Materials and Physics and the Thomas Young Centre for Theory and Simulation of Materials, Imperial College London, South Kensington Campus, London SW7 2AZ, UK\\}
\author{Xavier van Gorp}
\affiliation{Departments of Materials and Physics and the Thomas Young Centre for Theory and Simulation of Materials, Imperial College London, South Kensington Campus, London SW7 2AZ, UK\\}
\author{Dante M. Kennes}
\affiliation{Institute for Theory of Statistical Physics, RWTH Aachen University, and JARA Fundamentals of Future Information Technology, 52062 Aachen, Germany\\}
\affiliation{Max Planck Institute for the Structure and Dynamics of Matter, 
Center for Free Electron Laser Science, Luruper Chaussee 149, 22761 Hamburg, Germany\\}
\author{Arash A. Mostofi}
\affiliation{Departments of Materials and Physics and the Thomas Young Centre for Theory and Simulation of Materials, Imperial College London, South Kensington Campus, London SW7 2AZ, UK\\}
\author{Johannes Lischner}
\affiliation{Departments of Materials and Physics and the Thomas Young Centre for Theory and Simulation of Materials, Imperial College London, South Kensington Campus, London SW7 2AZ, UK\\}

\date{\today}

\begin{abstract}
Starting with twisted bilayer graphene, graphene-based moir\'e materials have recently been established as a new platform for studying strong electron correlations. In this paper, we study twisted graphene monolayers on trilayer graphene and demonstrate that this system can host flat bands when the twist angle is close to the magic-angle of 1.16$\degree$. When monolayer graphene is twisted on ABA trilayer graphene, the flat bands are not isolated, but are intersected by a Dirac cone with a large Fermi velocity. In contrast, graphene twisted on ABC trilayer graphene (denoted AtABC) exhibits a gap between flat and remote bands. Since ABC trilayer graphene and twisted bilayer graphene are known to host broken-symmetry phases, we further investigate the ostensibly similar magic angle AtABC system. We study the effect of electron-electron interactions in AtABC using both Hartree theory and an atomic Hubbard theory to calculate the magnetic phase diagram as a function of doping, twist angle, and perpendicular electric field. Our analysis reveals a rich variety of magnetic orderings, including ferromagnetism and ferrimagnetism, and demonstrates that a perpendicular electric field makes AtABC more susceptible to magnetic ordering.
\end{abstract}

\maketitle

\section{Introduction}

The observation of strong correlation phenomena in graphene-based moir\'e materials~\cite{Carr2020NatRev,TT,Tritsaris_2020,Balents2020,moiresim} has driven efforts to understand their electronic structure and behavior. A key prerequisite for the emergence of correlated states is flat electronic bands that give rise to a high density of states (DOS) at the Fermi level. The total energy of electrons in flat bands is dominated by the contribution from electron-electron interactions~\cite{MLWO,SMLWF,PHD_1}, which favors states that break symmetries of the Hamiltonian, opening gaps at the Fermi level to lower the total energy of the electrons. In moir\'e materials, it is possible to ``engineer" a high DOS at the Fermi energy through tuning the relative twist angle to values where very flat electronic bands emerge. 

In principle, the space of graphitic moir\'e systems is very large, but so far experimental studies have focused on five systems: twisted bilayer graphene (tBLG)~\cite{NAT_I,NAT_S,TSTBLG,SOM,Cao2020,SMTBLG,Polshyn2019,EFM,Serlin2020,Saito2020,Arora2020,Stepanov2020,Chern_Das,Chern_Wu,Chern_Nuckolls,NAT_MEI,NAT_SS,NAT_CO,IEC,Zondiner2020,Wong2020,Youngjoon2021}, twisted double bilayer graphene (tDBLG) comprosed of two AB stacked bilayers~\cite{BIBI,PhysRevLett.123.197702,RickhausPeter2019GOiT,TCT,cao2019electric,moireless2020,nematicity2020}, ABC trilayer graphene aligned with hexagonal boron nitride (ABC-hBN)~\cite{Chen2019ABC,Chen2019Mott,Chen2020ABC}, twisted mono-bilayer graphene (AtAB)~\cite{Yankowitz2020,Shi2020tTLG,Polshyn2020}, and twisted trilayer graphene (tTLG)~\cite{Tsai2019,hao2021engineering,cao2021large,Park2021}. Experimentally, all these systems have been found to exhibit correlated insulator states. Of particular interest are tBLG and tTLG (with an alternating twist angle between each sheet~\cite{hao2021engineering,cao2021large,Park2021}) because - in addition to correlated insulator states - robust superconductivity has been observed in these systems.

All of these systems have been predicted to feature flat electronic bands, which is a good indicator for possible strong correlations (tBLG~\cite{GBWT,MBTBLG,LDE,NSCS,MMIT,SCDID,KL,EPC,SCHFC,Bultinck2020,Zhang2020,Stauber2020,Cea2020,ECM,LK_CH,Ramires2019,Sboychakov2019,AF2020}, tDBLG~\cite{STSCTDB,ChoiY.W.2019Ibga,KoshinoMikito2019Bsat,ChebroluNarasimha2019Fbit,CulchacF.J.2020Fbag,WuFengcheng2020Fasi,Xia2020tDBLG}, AtAB~\cite{Park2020,Rademaker2020,MacDonald2019tTLG,Zhu2020}, tTLG~\cite{khalaf2019magic,Kruchkov2020,MacDonald2019tTLG,Zhu2020,Alejandro2020,Fischer_TTLG}, and ABC-hBN~\cite{Chen2019ABC,Chen2019Mott,Chen2020ABC}). In both tBLG~\cite{EE,Cea2019,Rademaker2019,PHD_4,Bascones2020} and tTLG~\cite{Fischer_TTLG} long-ranged electron-electron interactions lead to an additional enhancement of the DOS which increases the robustness of electronic correlations, and could be one reason why robust superconductivity is observed in these materials~\cite{Lewandowski2021,Cea2021}. In contrast, in tDBLG and mono-bilayer graphene, electric fields are required to further flatten the electronic bands and increase the DOS at the Fermi level~\cite{BIBI,PhysRevLett.123.197702,RickhausPeter2019GOiT,TCT,cao2019electric,Yankowitz2020,Shi2020tTLG,Polshyn2020}. Therefore, when investigating new graphitic moir\'e systems, it is important to investigate both the effect of electron-electron interactions on the band structure in the normal state and the response to external fields. 

\begin{figure*}[ht!]
\centering
\begin{minipage}[b]{0.32\textwidth}
    \centering 
    \includegraphics[width=\textwidth]{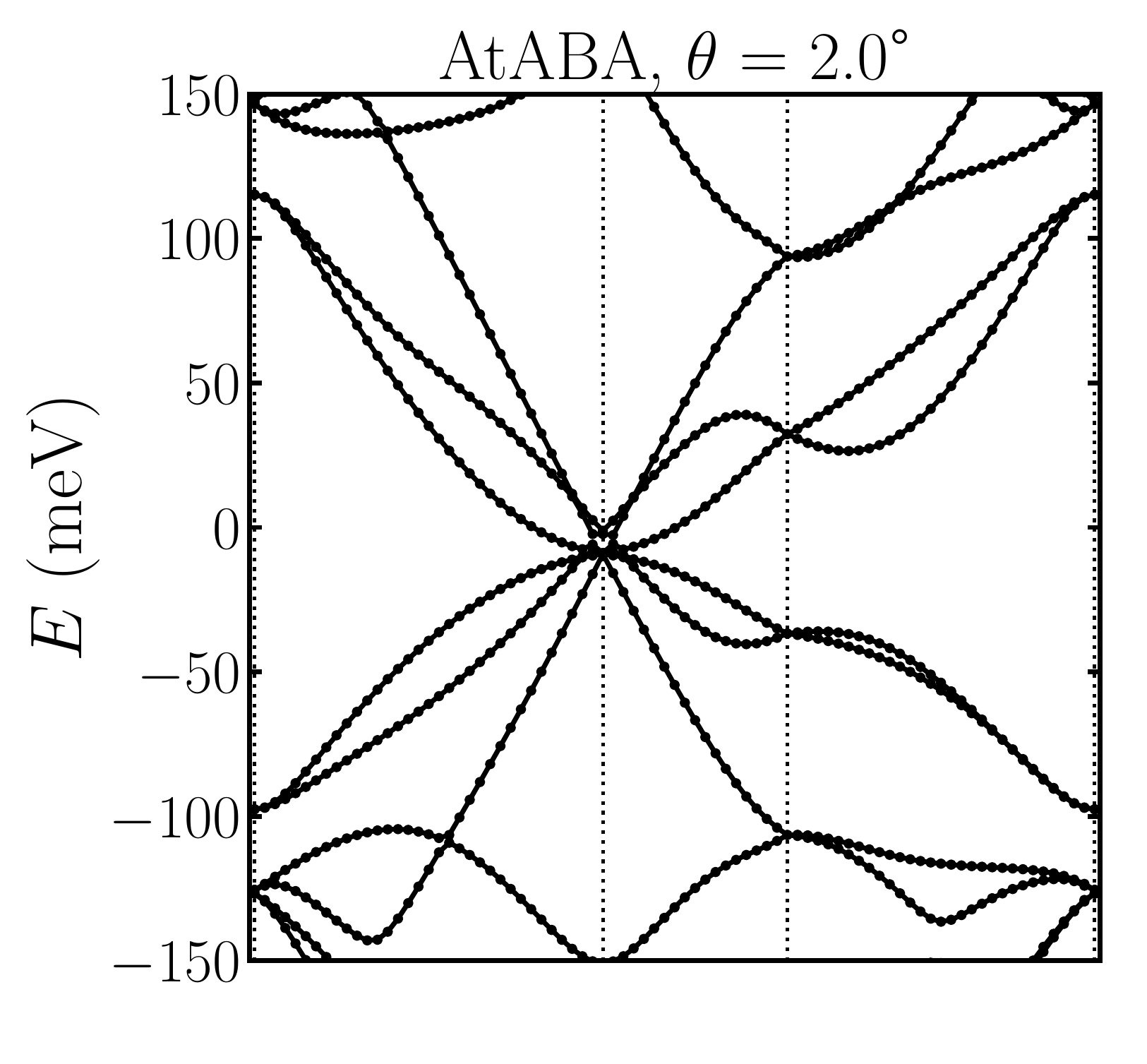}
\end{minipage}
\begin{minipage}[b]{0.32\textwidth}   
    \centering 
    \includegraphics[width=\textwidth]{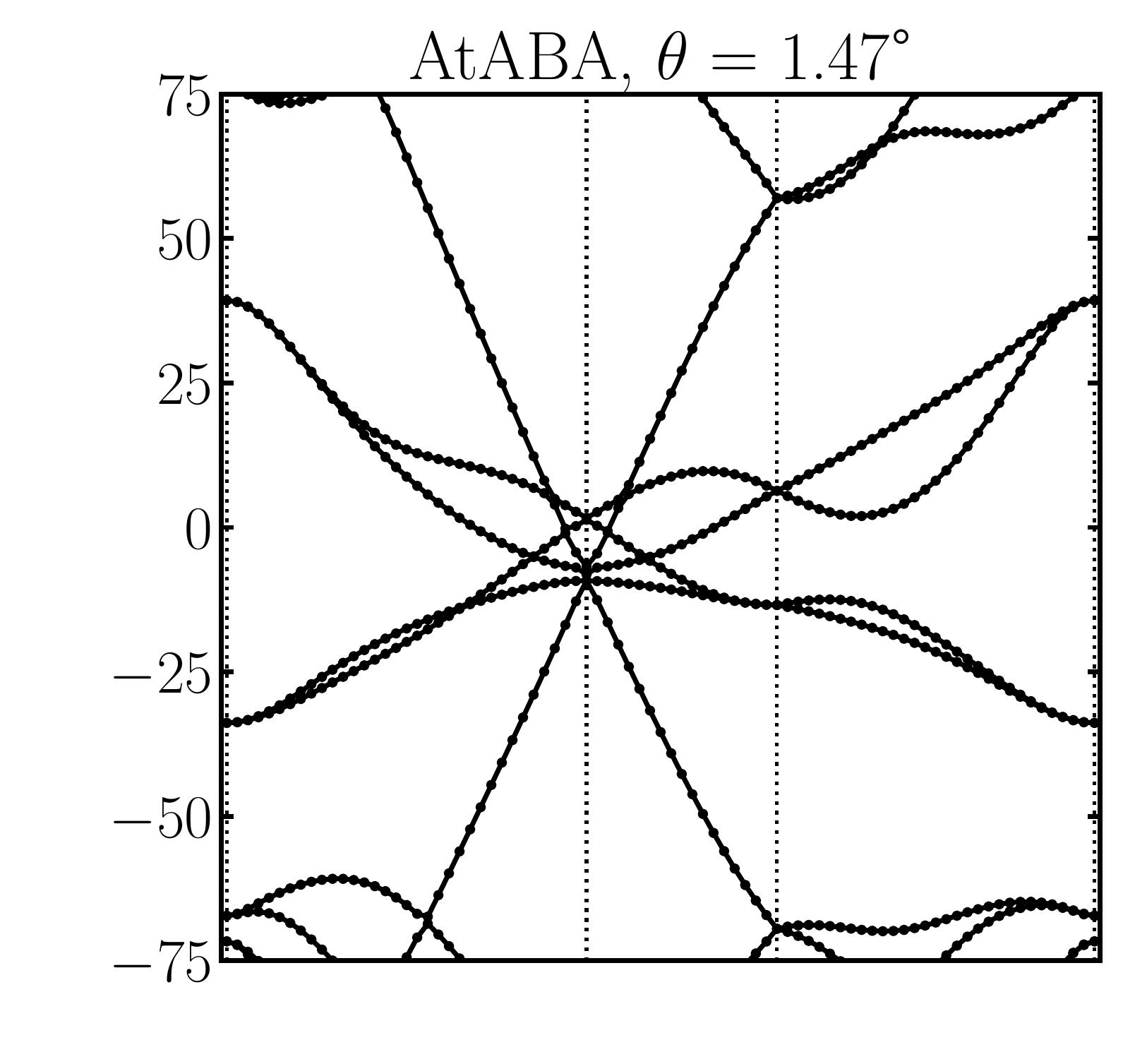}
\end{minipage}
\begin{minipage}[b]{0.32\textwidth}
    \centering 
    \includegraphics[width=\textwidth]{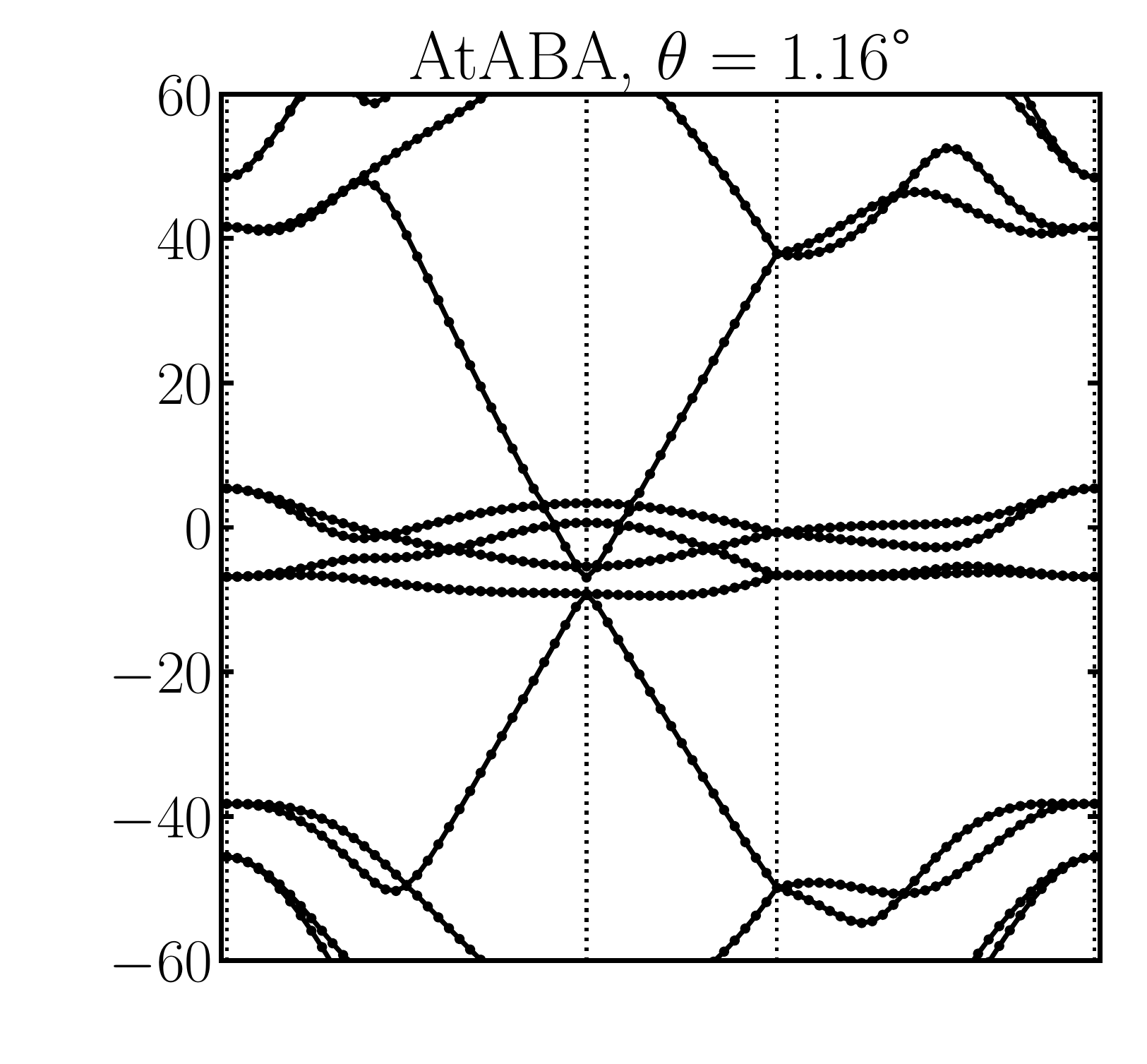}
\end{minipage}
\centering
\begin{minipage}[b]{0.32\textwidth}
    \centering 
    \includegraphics[width=\textwidth]{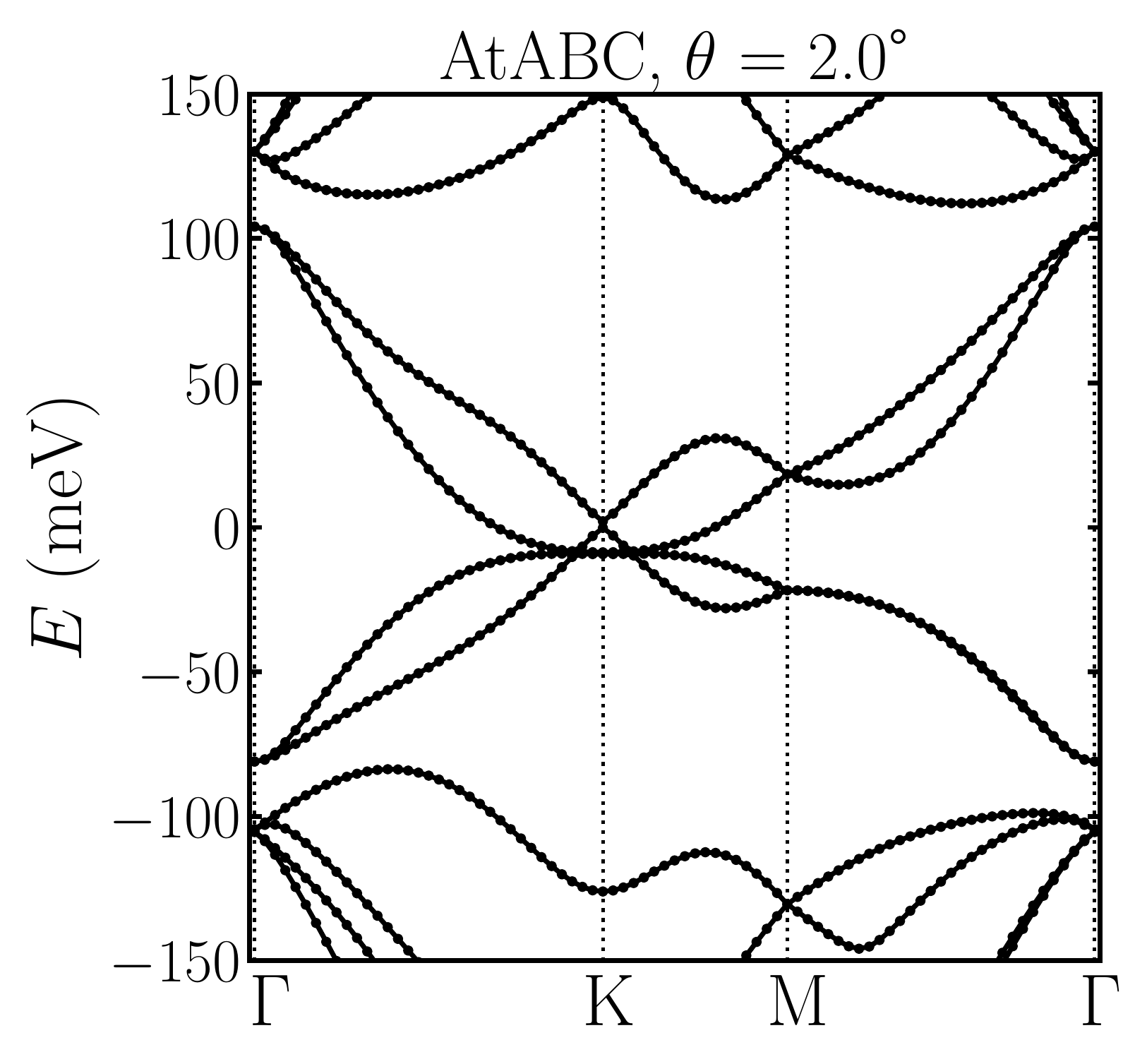}
\end{minipage}
\begin{minipage}[b]{0.32\textwidth}
    \centering 
    \includegraphics[width=\textwidth]{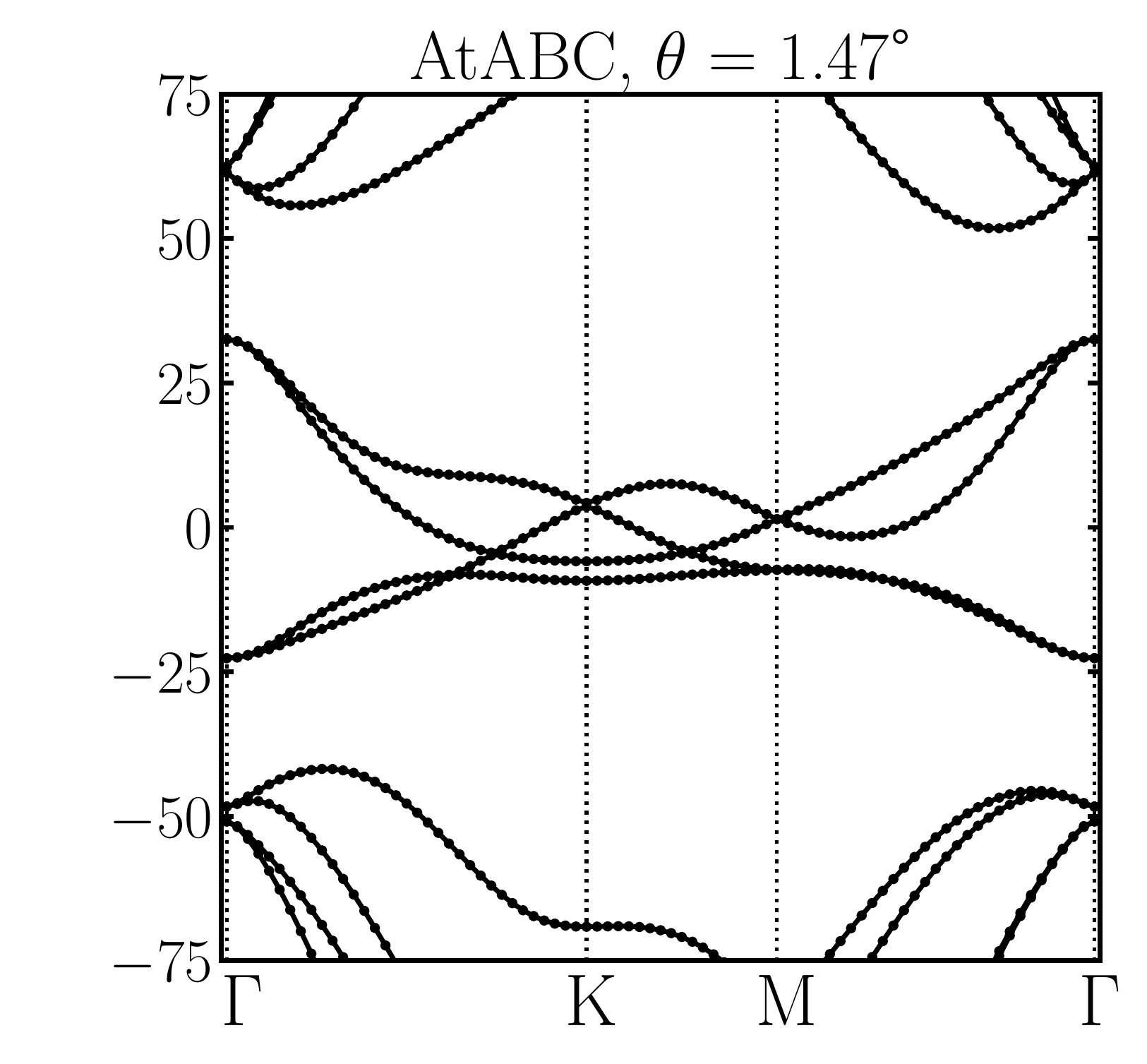}
\end{minipage}
\begin{minipage}[b]{0.32\textwidth}
    \centering 
    \includegraphics[width=\textwidth]{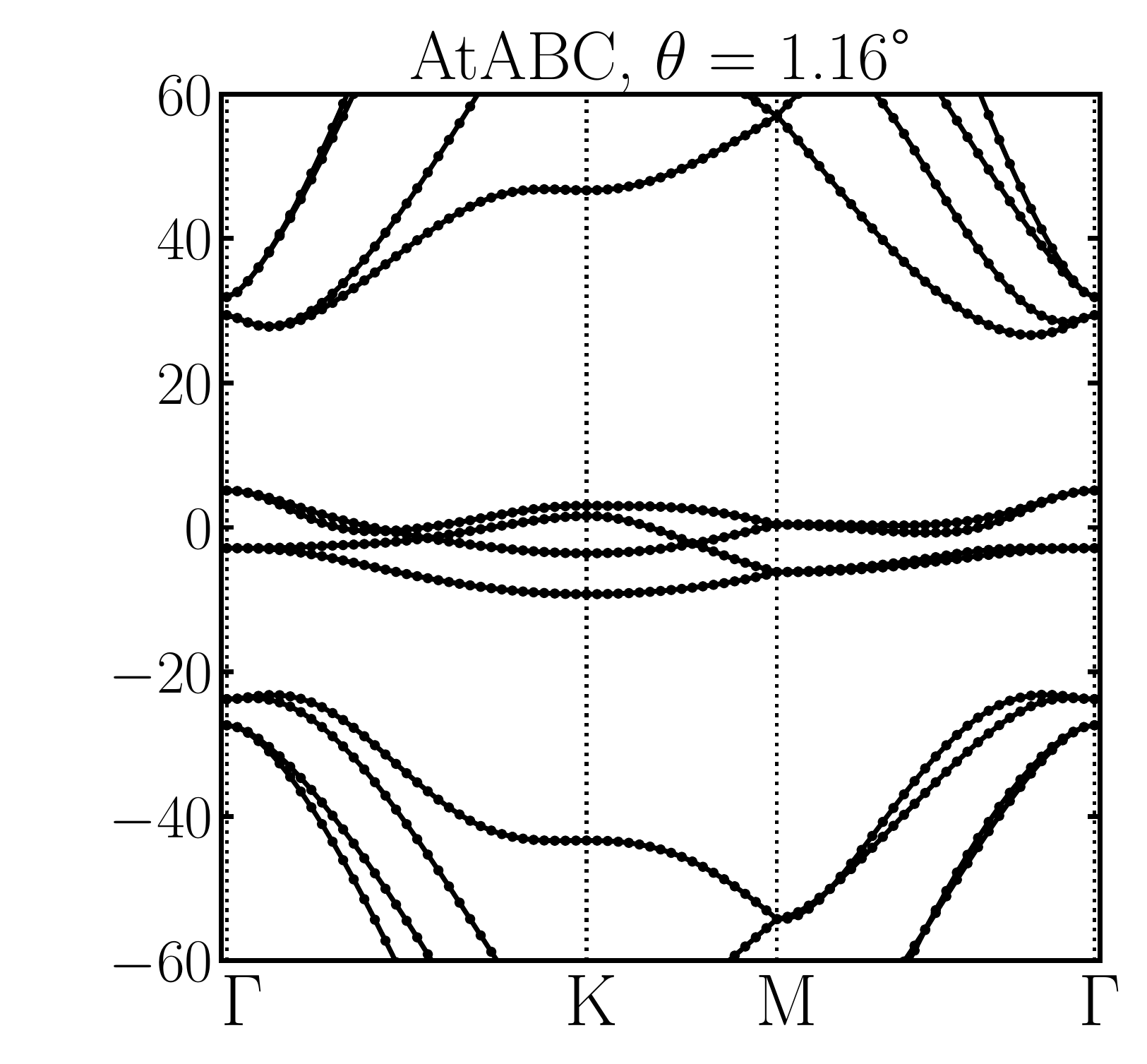}
\end{minipage}
\caption{Tight-binding band structure of AtABA and AtABC at three twist angles: 2.0$\degree$, 1.47$\degree$ and the magic-angle of 1.16$\degree$. Additional band structures for other twist angles are shown in Appendix B.} 
\label{fig:BS_0}
\end{figure*}

In this paper, we study the properties of twisted mono-trilayer graphene which consists of an untwisted graphene trilayer and a graphene monolayer that are twisted relative to each other. For the trilayer, we investigate both ABC and ABA stacking orders, which are both experimentally accessible. For twisted mono-ABC trilayer graphene (denoted AtABC, following the naming convention of Ref.~\citenum{Xian2020}), a set of four isolated flat bands emerges at the Fermi level, while in twisted mono-ABA trilayer graphene (denoted AtABA) the four flat bands are not isolated, but are intersected by a Dirac cone with a large Fermi velocity. In contrast to tBLG~\cite{EE,Cea2019,Rademaker2019,PHD_4,Bascones2020} and tTLG~\cite{Fischer_TTLG}, long-ranged Hartree interactions have little effect on the band structure. We find that short-ranged Hubbard interactions give rise to a rich magnetic phase diagram as a function of twist angle, doping level and perpendicular electric field that features competing anti-ferromagnetic and ferrimagnetic orderings.

\begin{figure}
\centering
\includegraphics[width=0.45\textwidth]{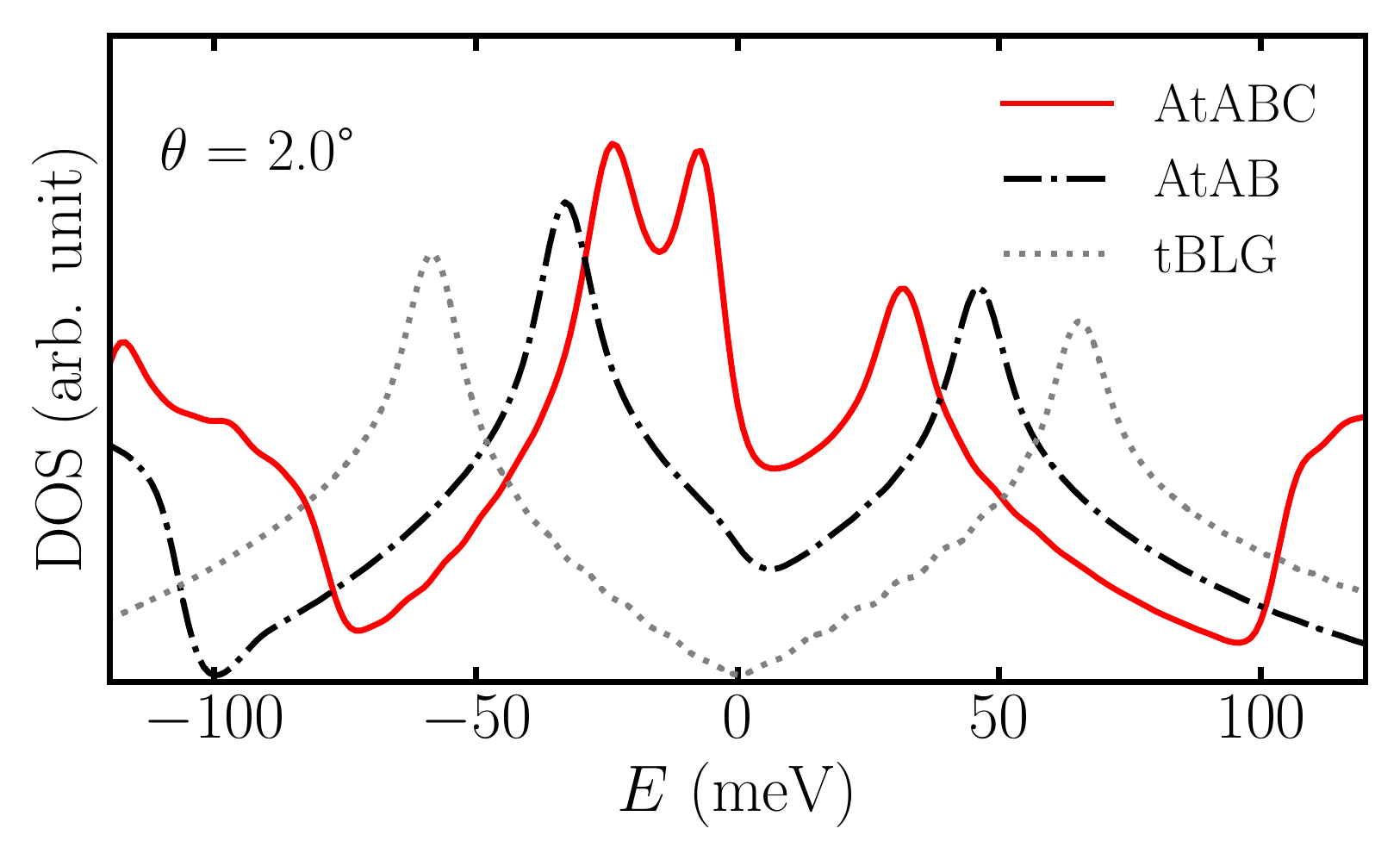}
\caption{Density of states (DOS) as a function of energy for AtABC, AtAB and tBLG at a twist angle of 2.0$\degree$. The zero of energy is set to the Fermi level at charge neutrality for each system, and the zero in the DOS is at the bottom of the $y$-axis.}
\label{fig:DOS}
\end{figure}

\section{Results and Discussion}

\subsection{Atomic structure}

As AtABC and AtABA only contain a single twist angle, the approach of Ref.~\citenum{NSCS}, initially developed for tBLG, can be used to generate commensurate moir\'e unit cells for mono-trilayer systems (see Appendix A for details). We relax these structures using classical force fields to determine the equilibrium positions of the atoms. Specifically, we employ the adaptive intermolecular reactive empirical bond order-Morse(AIREBO-Morse)~\cite{AIREBO} potential for intralayer interactions and the Kolmogorov-Crespi~\cite{KC} potential for interlayer interactions, as implemented in the Large-Scale Atomic/Molecular Massively Parallel Simulator (LAMMPS)~\cite{LAMMPS}. Further details can be found in Appendix A and in Ref.~\citenum{Xia2020tDBLG}. 

\subsection{Electronic structure}

We first calculate the electronic band structure of AtABC and AtABA at different twist angles using an atomistic tight-binding approach, see Fig.~\ref{fig:BS_0} and Appendix B for details of the calculation. For both systems, we find that a set of extremely flat electronic bands emerges as the twist angle approaches the magic-angle of 1.16$\degree$~\footnote{Note that the exact value of magic angle depends on the choice of hopping parameters in the tight-binding model. For tBLG, the hopping parameters used in our model result in a magic angle of 1.05$\degree$ which is 0.05$\degree$ smaller than the currently accepted value extracted from experiments~\cite{Balents2020}. We therefore expect a similar accuracy for the structures studied in this work.}. 

At twist angles larger than the magic angle (2.0$\degree$ and 1.47$\degree$), the band structure of AtABA [in Fig.~\ref{fig:BS_0} (top panels)] exhibits a set of four bands with a bandwidth of the order of 100~meV. Two of these form a Dirac cone at the K-point while the other two have a parabolic dispersion near K. At the magic angle of 1.16$\degree$ (and also at smaller twist angles) the bands no longer form a Dirac cone. The low-energy bands in AtABA are not isolated in energy from the remote bands because they are intersected by a pair of linear bands whose Fermi velocity is similar to that of monolayer graphene. These bands form a second Dirac cone at K which exhibits a small gap of $\sim 1-5$~meV.

Additional insight can be gained by comparing the band structure of AtABA with that of the constituent ABA trilayer. The latter system features a set of parabolic bands which are also intersected by a Dirac cone~\cite{Menezes2014}. This suggests that the addition of the twisted graphene monolayer on top of the ABA trilayer induces the ``flat" Dirac cone (whose Dirac point lies slightly higher in energy than that of the dispersive Dirac cone) and also modifies the bandwidth of the parabolic bands. Finally, it is also interesting to note that the band structure of AtABA is quite similar to that of twisted trilayer graphene in which the middle layer of an AAA-stacked trilayer is twisted relative to the outer layers~\cite{khalaf2019magic,Kruchkov2020,MacDonald2019tTLG,Zhu2020,Alejandro2020,Fischer_TTLG}.

%
%
\begin{figure*}[ht]
\centering
\begin{minipage}[b]{0.49\textwidth}
    \makebox[8pt]{\footnotesize\bfseries(a)}
    \includegraphics[valign=t,width=0.9\textwidth]{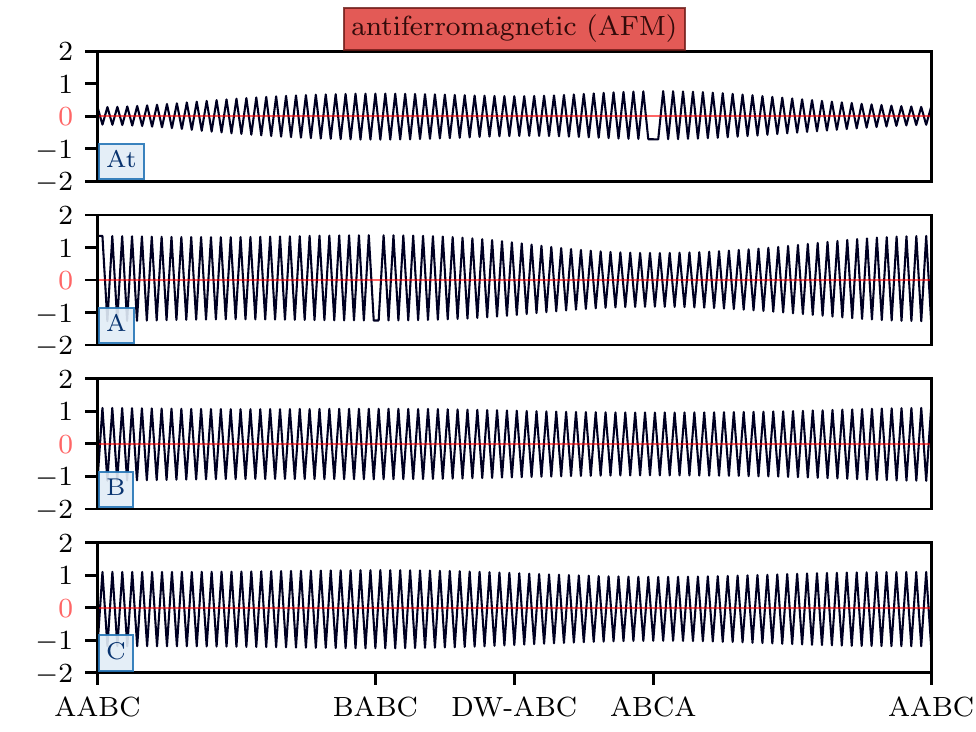}
\end{minipage}
\begin{minipage}[b]{0.49\textwidth}
    \makebox[8pt]{\footnotesize\bfseries(b)}
    \includegraphics[valign=t,width=0.9\textwidth]{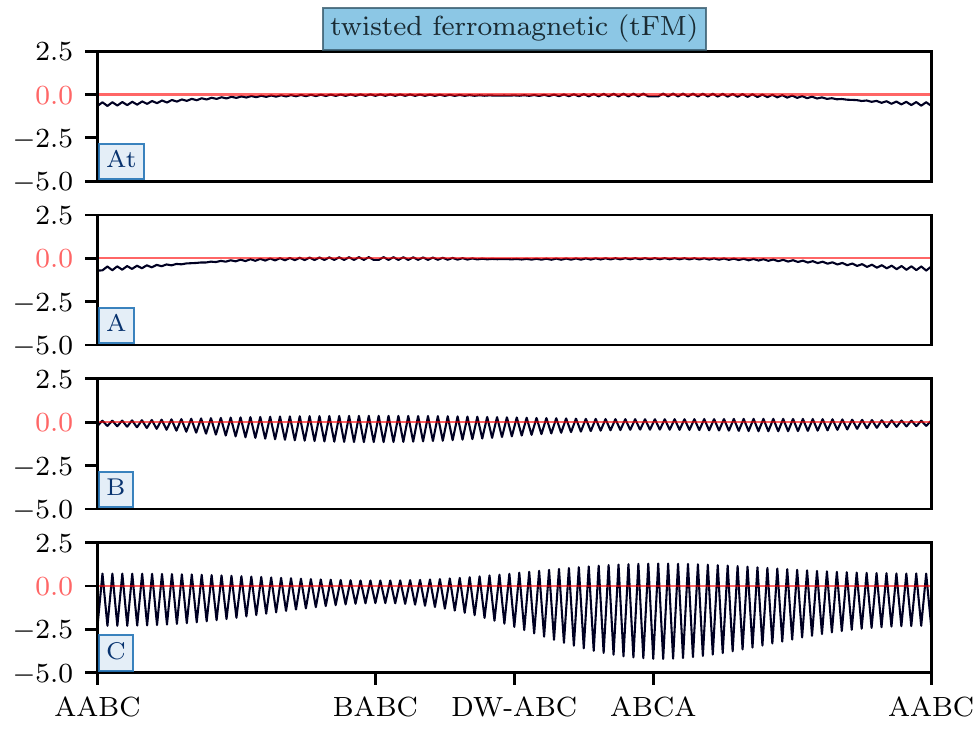}
\end{minipage} \\[5pt]
\begin{minipage}[b]{0.49\textwidth}
    \makebox[8pt]{\footnotesize\bfseries(c)}
    \includegraphics[valign=t,width=0.9\textwidth]{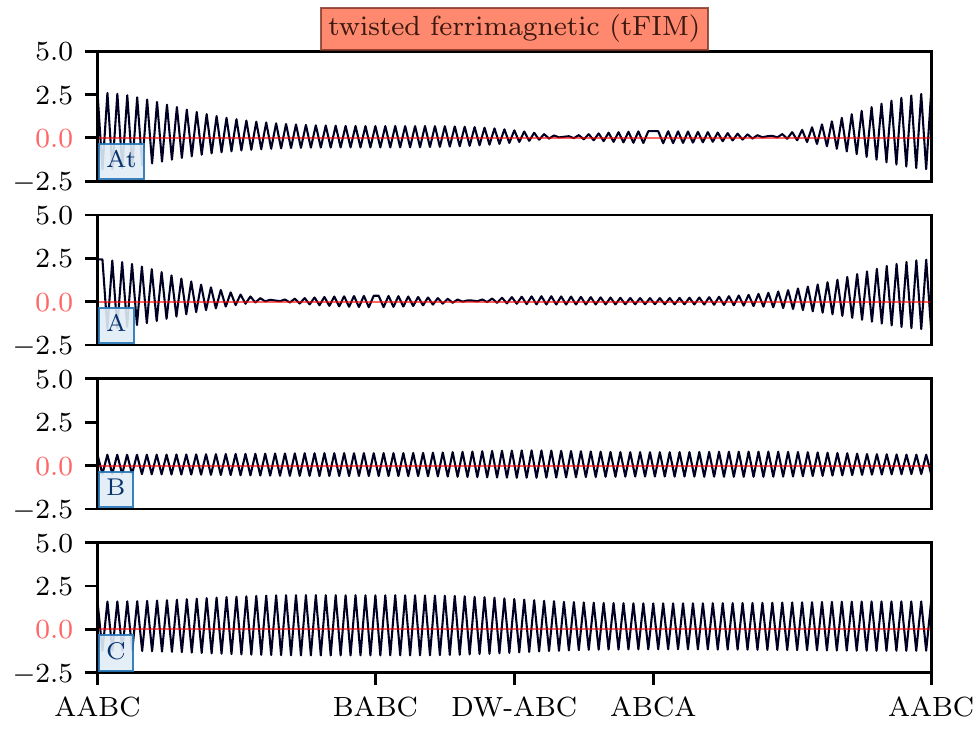}
\end{minipage}
\begin{minipage}[b]{0.49\textwidth}
    \makebox[8pt]{\footnotesize\bfseries(d)}
    \includegraphics[valign=t,width=0.9\textwidth]{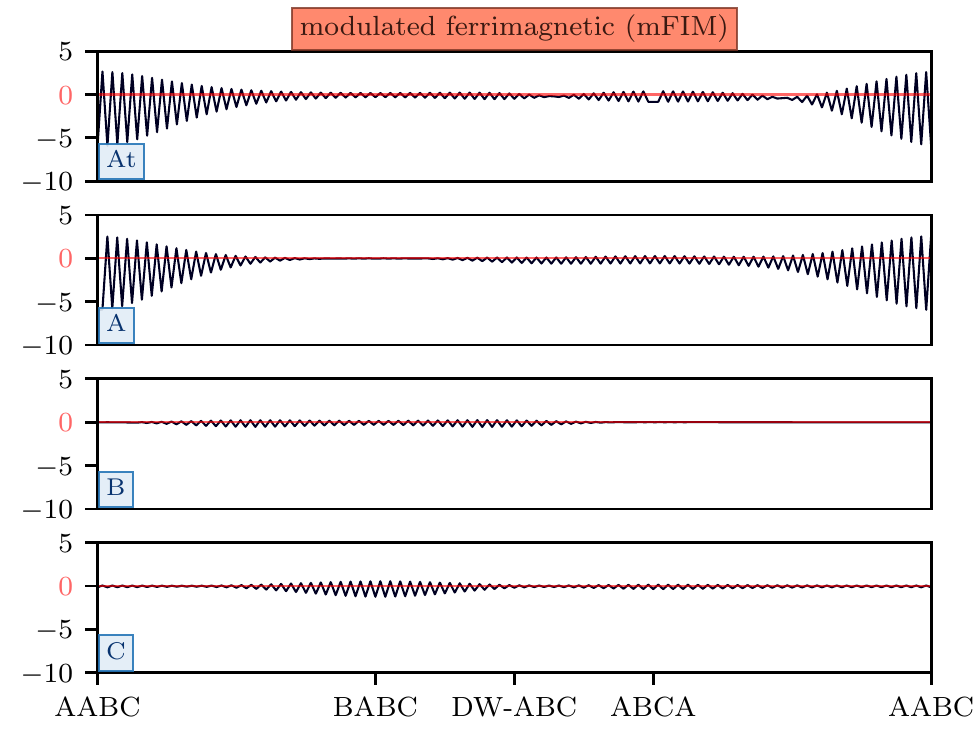}
\end{minipage}
\caption{
Layer-resolved line-cuts of the magnetization for different magnetic instabilities (the $y$-axis is the normalized eigenvector corresponding to eigenvalue $\chi^0$) of AtABC along the diagonal of the moir\'e unit cell at the magic angle of $\theta=1.16\degree$. The local stacking sequence is shown at the bottom of each panel, where DW stands for the domain wall region of the moir\'e pattern. {\bfseries{}(a)} Anti-ferromagnetic state with mild modulations on the moir\'e scale. {\bfseries{}(b)} A state with ferromagnetic order in the twisted layers and ferrimagnetic order in the lower layers. {\bfseries{}(c)} A state with modulated ferrimagnetic order in the twisted layers and relatively uniform anti-ferromagnetic order in the lower layers. {\bfseries{}(d)} A state with modulated ferrimagnetic order in all layers.}
\label{fig:linecut}
\end{figure*}

Figure~\ref{fig:BS_0} (bottom panels) also shows the band structure of AtABC as a function of twist angle. For this system we also find a set of four flat bands near the Fermi level. Although these bands look qualitatively similar to those of AtABA, there are some important differences. As in AtABA, the low-energy electronic structure of AtABC has one pair of bands that form a Dirac cone at K for twist angles larger than the magic angle. The other pair of bands, however, now has a cubic dispersion near K, and there is no additional Dirac cone that intersects these bands, which are entirely separated from the remote bands in this system near the magic angle. 

Again, it is instructive to compare the band structure of AtABC with that of the constituent parts. In ABC trilayer graphene, there is a set of cubic bands near the Fermi level~\cite{Min2008}, which AtABC retains in the the low-energy dispersion of the isolated bands with the twist angle controlling their width. Finally, it is worth noting that the band structure of AtABC is similar to that of twisted monolayer-AB bilayer graphene (AtAB), with the important difference that the dispersion in AtAB is parabolic~\cite{Shi2020tTLG} instead of cubic at the K point, as shown in Appendix B. 

This difference in the power law of the dispersion has important consequences for the DOS. In Fig.~\ref{fig:DOS}, we show the DOS of the flat bands of AtABC, AtAB and tBLG at an angle of 2.0$\degree$ (see Appendix C for further twist angles). All systems have a pair of van Hove singularities at an energy corresponding to a doping level of $\pm2$ electrons (relative to charge neutrality) per moir\'e unit cell. The linear dispersion of the tBLG bands close to charge neutrality gives rise to a linear DOS close to the Dirac point where the DOS vanishes. In contrast, for AtAB, the DOS is always finite and exhibits a step-like feature at approximately $-$5~meV where the parabolic bands touch. Importantly, the AtABC system has an additional van Hove singularity arising from the bands with a cubic dispersion. 

\subsection{Electron-electron interactions}

Based on the tight-binding calculations, we have identified AtABC as a promising candidate for hosting strongly correlated electrons in isolated flat bands. We therefore study the effect of electron-electron interactions in this system. To capture the effect of long-ranged Coulomb interactions, we carry out self-consistent atomistic Hartree theory calculations at integer doping levels per moir\'e unit cell. However, in contrast to tBLG and tTLG, we find that such interactions have a negligible effect on the electronic band structure of AtABC, see Appendix D. 

This can be understood by analyzing the spatial character of the wavefunctions at different points in the first Brillouin zone. In tBLG~\cite{Rademaker2020} and tTLG~\cite{Fischer_TTLG}, it was shown that there is a strong correlation between a state's position in k-space and its localization in real space. For example, the states at the edge of the hexagonal Brillouin zone are localized on the AA regions while states at the center of the Brillouin zone are localized on the AB regions. When the occupancy of these states changes due to electron or hole doping a highly inhomogeneous charge density is induced, which in turn results in a strong Hartree potential. In contrast, we do not observe a similar correlation between k-space position and real-space localization of states in AtABC and as a consequence the charge density induced by doping is relatively uniform resulting in a much weaker Hartree potential. This result is consistent with other Hartree calculations of moir\'e materials containing untwisted graphene layers~\cite{Pierre2020}.

In the absence of significant Hartree interactions, we next consider the effect of exchange interactions. It is well known that the exchange interaction should be screened~\cite{AshcroftMermin} which reduces its strength and modifies its spatial form. In moir\'e materials, the presence of flat bands greatly enhances the internal screening~\cite{PHD_2,CCRPA}, and external screening arising from the presence of nearby metallic gates further suppresses long-ranged interactions~\cite{PHD_3}. As a consequence of screening, the range of the exchange interaction is significantly shorter than the moir\'e length scale and we therefore employ an atomic Hubbard interaction for electrons in the carbon p$_z$-orbitals~\cite{ECM,LK_CH,PHD_6,Ramires2019,Sboychakov2019,AF2020,Fischer_TTLG} to calculate the interacting spin susceptibility using the random-phase approximation (RPA) as a function of doping, twist angle and value of the Hubbard $U$ parameter. This approach was also employed in Ref.~\citenum{PHD_6} for tBLG, where excellent agreement between the experimental and theoretical phase diagram was found, and functional renormalization group calculations~\cite{LK_DMK_CH} have shown that the phase diagram is not sensitive to the range of interactions provided it is short. Therefore, we are confident that this approach can reliably identify the onset of broken symmetry phases in graphitic moir\'e systems.

From these RPA calculations, we identify the critical value of the Hubbard parameter $\Ucrit$ at which the susceptibility diverges~\cite{LK_CH,PHD_6}. If $\Ucrit$ is smaller than the physical value of the Hubbard parameter, we expect the system to undergo a phase transition into a magnetically ordered state whose spatial structure is determined by the leading eigenvector of the spin response function. In this paper, we use a Hubbard value of $U=4$~eV, which has been shown to be a realistic value of the onsite Hubbard interaction of graphene~\cite{SECI,OHP}. Moreover, in Ref.~\citenum{PHD_6}, it was shown that $U=4$~eV for tBLG yields good agreement with the available experimental data. For additional details about the method, see Appendix E and Ref.~\citenum{LK_CH}.

\begin{figure*}
\centering
\null \hfill 
\makebox[12pt]{{\bf(a)}}
\includegraphics[valign=t,scale=0.9]{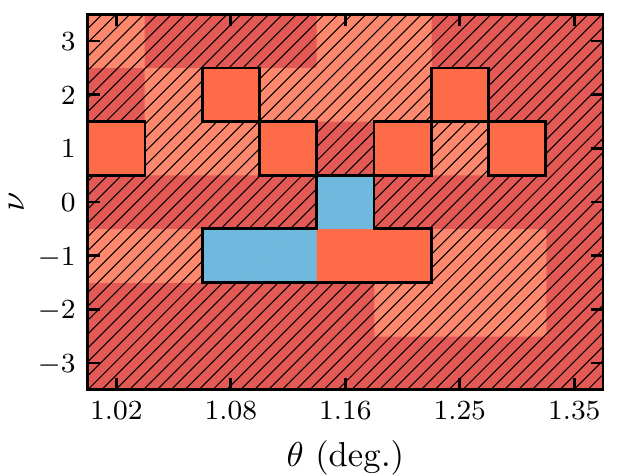}
\hfill 
\makebox[12pt]{{\bf(b)}}
\includegraphics[valign=t,scale=0.9]{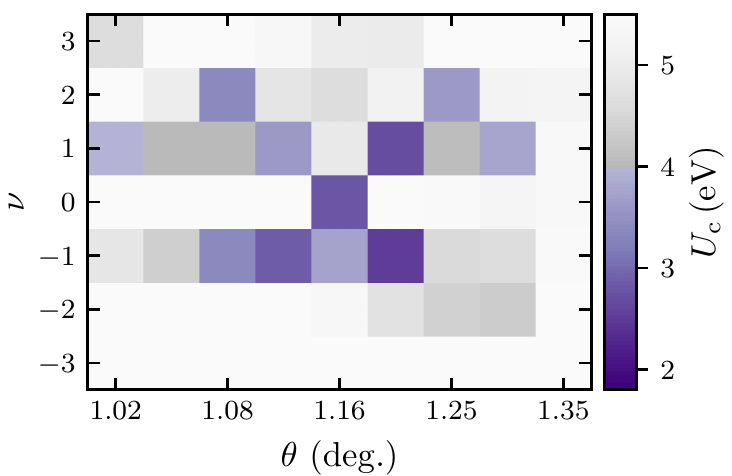}
\hfill \null
\caption{{\bfseries{}(a)} Magnetic phase diagram of AtABC as a function of twist angle $\theta$ and integer doping level $\nu$ per moir\'e unit cell. Blue corresponds to ferromagnetic order, orange corresponds to ferrimagnetic order, and red corresponds to anti-ferromagnetic order, examples of which can be seen in Fig.~\ref{fig:linecut}. Regions where the critical value of the Hubbard parameter is smaller than its physical value are hatched. {\bfseries{}(b)} Critical interaction strength $\Ucrit$ required for the onset of magnetic instabilities in AtABC as a function of twist angle $\theta$ and integer doping levels in the flat bands $\nu$.}
\label{fig:twistangle}
\end{figure*}

Figure~\ref{fig:linecut} shows the structure of various low-energy magnetic states of AtABC at the magic-angle of 1.16$\degree$. In each of the plots, we display a normalized eigenvector from the magnetic susceptibility calculations as a function of position along the diagonal of the moir\'e unit cell (different stacking regions are indicated on the $x$-axis). Other leading instabilities were also found, but they are either variations of the ones shown in Fig.~\ref{fig:linecut} with a different nodal structure or mixtures of these orderings. Overall, we find that there is a rich variety of magnetic ordering tendencies that can be dominant either in the twisted layers or in the untwisted layers.

Figure~\ref{fig:linecut}(a) shows an antiferromagnetic (AFM) state which is mostly uniform over the whole AtABC structure and only exhibits a mild modulation on the moir\'e scale. The magnetization differs slightly in each layer with the largest variations occurring in the graphene sheet that is twisted on top of the ABC trilayer. This layer also exhibits a smaller magnitude of the magnetization than in the other layers, suggesting that this AFM state is inherited from the AFM state of the ABC trilayer which ``spills" into the top layer.

Figure~\ref{fig:linecut}(b) shows a state with a modulated ferromagnetic (FM) structure in the top two layers and ferrimagnetic structure in the bottom two layers. We refer to this ordering as tFM (for twisted FM, as the FM order is found in the twisted layers). In the upper layers the magnetization has peaks in the AABC regions which are separated by a node. A similar state has been found in tBLG~\cite{LK_CH,PHD_6}. Finally, Figs.~\ref{fig:linecut}(c) and (d) show two examples of ferrimagnetic (FIM) states. The state in Fig.~\ref{fig:linecut}(c) is mostly AFM with some FIM character and exhibits nodes in the top two layers. We shall refer to this ordering as tFIM (for twisted FIM, as the FIM order is mainly in the twisted layers). The state in Fig.~\ref{fig:linecut}(d) is predominantly FIM and has significant modulations in each layer, which we refer to as mFIM (for modulated FIM).

Having described in detail the different types of magnetic states in AtABC, we now discuss the magnetic phase diagram as a function of twist angle and doping, denoted by $\nu$ for the number of additional electrons or holes per moir\'e unit cell, as shown in Fig.~\ref{fig:twistangle}(a). Magnetic states with $\Ucrit < U=4$~eV are found for a range of doping levels and twist angles. When $\Ucrit > U=4$~eV, we hatch over the magnetic order to indicate that we do not expect it to occur. In Fig.~\ref{fig:twistangle}(b) we plot the corresponding value of $\Ucrit$ for each $\theta-\nu$ combination and if $\Ucrit > U=4$~eV we use a gray scale. The FM state is only found at the magic angle at charge neutrality or at slightly smaller twist angles for $\nu=-1$ (i.e., when one hole is added per moir\'e unit cell). Interestingly, the character of the FM state for $\nu=-1$ slowly transitions from purely FM to a mixture of FM and FIM as a function of the twist angle.

The other broken-symmetry states in the phase diagram are of FIM type and occur not only at $\nu=-1$ at and very close to the magic-angle, but also for the electron doped systems ($\nu=1$ or $\nu=2$) over a range of twist angles. In contrast, AFM order is never found in the phase diagram. While this type of order is the leading instability for a range of $\theta-\nu$ values, the corresponding critical values of the Hubbard parameter are always larger than the physical value ($\Ucrit > U=4\,\mathrm{eV}$) and therefore this order is not realized. This can be attributed to the fact that the AFM order is inherited from the parent ABC trilayer system which has a high value of $\Ucrit$~\cite{Scherer2012}. 

\begin{figure*}
\centering
\null \hfill %
\makebox[12pt]{{\bf(a)}}
\includegraphics[valign=t,scale=0.9]{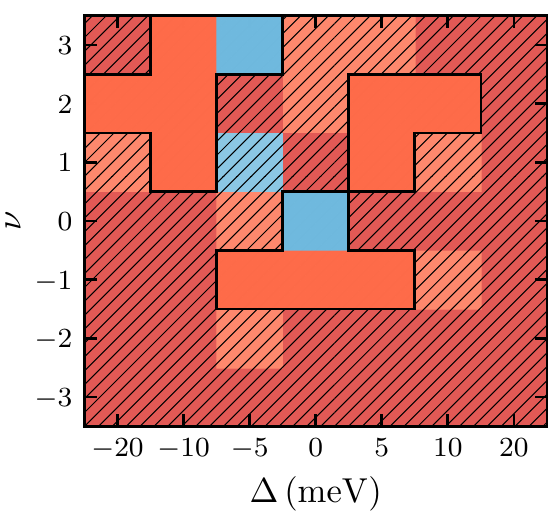}
\hfill %
\makebox[12pt]{{\bf(b)}}
\includegraphics[valign=t,scale=0.9]{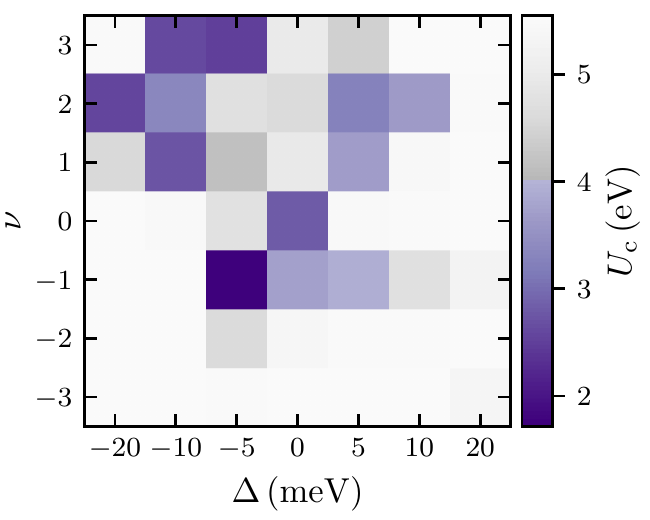}
\hfill \null
\caption{{\bfseries{}(a)} Magnetic phase diagram of 1.16$\degree$ AtABC as a function of potential difference between adjacent layers $\Delta$ (which is proportional to a perpendicular electric field) and integer doping level $\nu$ per moir\'e unit cell. Blue corresponds to ferromagnetic order, orange corresponds to ferrimagnetic order, and red corresponds to anti-ferromagnetic order, examples of which can be seen in Fig.~\ref{fig:linecut}. Regions where the critical value of the Hubbard parameter is smaller than its physical value are hatched. {\bfseries{}(b)} Critical interaction strength $\Ucrit$ required for the onset of magnetic instabilities in 1.16$\degree$ AtABC as a function of layer-dependent onsite potential difference $\Delta$ and integer doping levels in the flat bands $\nu$.}
\label{fig:efield}
\end{figure*}

Having presented the band structure, effects of electron-electron interactions and magnetic order of AtABC, a natural question to ask is: how promising is AtABC for the observation of strong correlation phenomena in comparison to other graphitic moir\'e materials? Among the graphene-based moir\'e materials that have been studied experimentally to date, only tBLG and tTLG exhibit robust superconductivity~\cite{NAT_S,Park2021}. In contrast to AtABC, the long-ranged Coulomb interaction plays an important role in these systems and enlarges the size of the region in the $\theta-\nu$ phase diagram where broken symmetry states occur~\cite{PHD_6,Lewandowski2021}. Based on this empirical evidence, one could argue that moir\'e systems that do not contain any untwisted pairs of neighboring layers~\cite{khalaf2019magic} are more promising candidates for the observation of strongly correlated phases than moir\'e materials that contain untwisted layers~\cite{Choi2021Dichotomy}. While this might be true in the absence of electric fields, recent reports suggest that magic-angle mono-bilayer (AtAB) graphene exhibits both correlated insulating states~\cite{Yankowitz2020} and signatures of superconductivity when an electric field is applied perpendicular to the layers~\cite{Shi2020tTLG}. For comparison against AtABC, we have also calculated the phase diagram of AtAB, see Appendix E. Our analysis reveals that these systems exhibit qualitatively similar types of magnetic order, which suggests that AtABC may also be a promising candidate for the observation of strong correlation phenomena in the presence of applied electric fields. 

To put this prediction on a stronger footing, we calculated the interacting spin susceptibility of magic-angle (1.16$\degree$) AtABC as a function of applied electric field and doping, as shown in Fig.~\ref{fig:efield}. A perpendicular electric field introduces an additional onsite potential that is approximately constant within a layer, but varies linearly between the layers. We define $\Delta$ as the potential difference between two adjacent layers, such that the onsite potential of layer $l$ (where $l=1,2,3,4$ with $1$ corresponding to the twisted monolayer) is given by $-\Delta\cdot l$. Negative values of $\Delta$ mean that the potential energy of the electrons is lowest in the twisted monolayer. This potential difference is directly proportional to the applied electric field, with values of $|\Delta| = 30$~meV being well within experimental reach.

In the absence of a field, we only expect magnetic order to occur at charge neutrality or $\nu=-1$ at 1.16$\degree$ [see Fig.~\ref{fig:twistangle}(a)]. Upon applying an electric field which lowers the energy of electrons in the twisted layers ($\Delta < 0$), we find that the system is more susceptible to magnetic ordering. Overall, we find mainly FIM order in electron-doped systems, but the hole-doped systems do not generally become more susceptible to magnetic ordering, with the exception of $\nu=-1$ at $\Delta = -5$~meV [see Fig.~\ref{fig:efield}(b)]. Therefore, the electron-hole asymmetry of the magnetic phase diagram becomes more pronounced in an electric field which lowers the energy of the electrons in the twisted layers relative to the other layers. On the other hand, electric fields which increase the energy of the electrons in the twisted layers ($\Delta > 0$) generally cause the system to be less susceptible to magnetic ordering. For electron-doped $\Delta > 0$ systems in a small field, we find that FIM occurs at $\nu=1,2$, but for larger field strengths this magnetic order disappears. In experiments on magic-angle AtAB performed by Chen \textit{et al.}~\cite{Yankowitz2020}, there were similar trends in terms of where the correlated insulating states occur in the space of doping level and electric field. For an electric field which lowers the energy of the monolayer (relative to the AB bilayer), correlated insulating states were found at all integer electron doping levels, similar to tBLG~\cite{Yankowitz2020}. Whereas, for an electric field which lowers the energy of the AB stacked bilayer (relative to the monolayer), a correlated insulating state was only observed at $\nu=2$, similar to tDBLG~\cite{Yankowitz2020}. As we have found that AtABC has a similar electronic structure and electron interactions to AtAB, this also suggests similarities in their broken symmetry phases.

In summary, we have established magic-angle AtABC as a highly promising candidate for the observation of broken symmetry phases, such as magnetic order. To test our predictions, we propose that transport experiments on magic-angle AtABC should be carried out to determine the phase diagram as a function of doping, with Hall measurements being able to discern if the phase has ferromagnetic character. Additionally, scanning tunneling microscopy can be used to verify the presence of an additional van Hove singularity in AtABC (compared to tBLG or AtAB) and also the predicted weakness of Hartree interactions. These measurement techniques can also identify correlated insulating states and superconductivity, if present. Promising future directions for theoretical work on AtABC are the study of its topological properties, possible superconductivity mechanisms, nematic ordering, and cascade instabilities.

\begin{figure*}
\centering
\begin{minipage}[b]{0.45\textwidth}
\centering
\includegraphics[width=\textwidth]{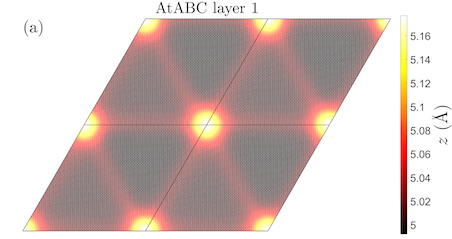}
\end{minipage}
\begin{minipage}[b]{0.45\textwidth}  
\centering 
\includegraphics[width=\textwidth]{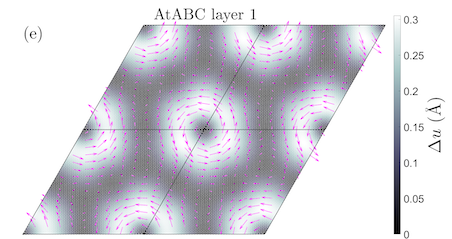}
\end{minipage}
\hfill
\vskip\baselineskip
\begin{minipage}[b]{0.45\textwidth}  
\centering 
\includegraphics[width=\textwidth]{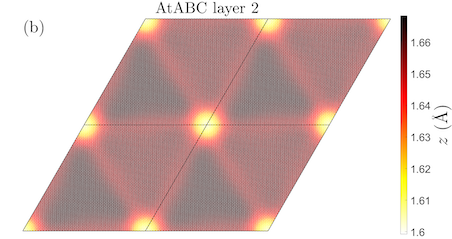}
\end{minipage}
\begin{minipage}[b]{0.45\textwidth}   
\centering 
\includegraphics[width=\textwidth]{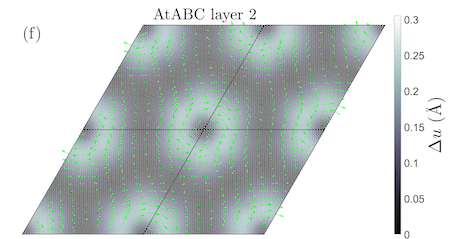}
\end{minipage}
\vskip\baselineskip
\begin{minipage}[b]{0.45\textwidth}
\centering
\includegraphics[width=\textwidth]{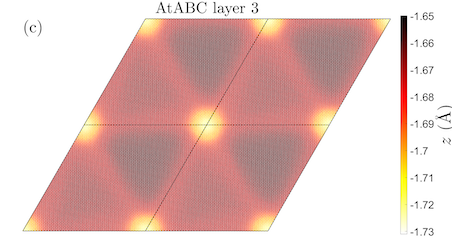}
\end{minipage}
\begin{minipage}[b]{0.45\textwidth}  
\centering 
\includegraphics[width=\textwidth]{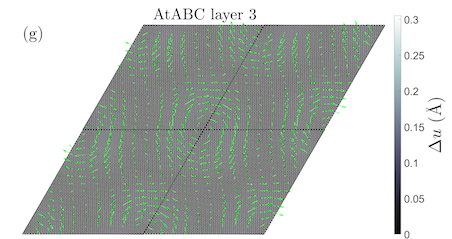}
\end{minipage}
\hfill
\vskip\baselineskip
\begin{minipage}[b]{0.45\textwidth}  
\centering 
\includegraphics[width=\textwidth]{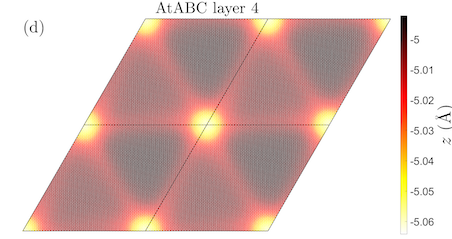}
\end{minipage}
\begin{minipage}[b]{0.45\textwidth}   
\centering 
\includegraphics[width=\textwidth]{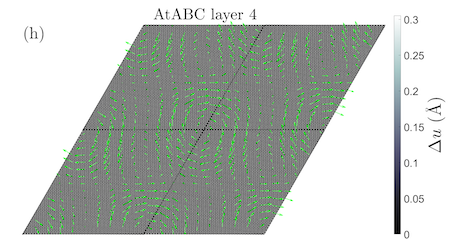}
\end{minipage}
\caption{Out-of-plane and in-plane relaxations of AtABC for a twist angle of $\theta=0.73\degree$. [(a)-(d)] Out-of-plane displacements for layers 1 to 4, respectively; [(e)-(h)] In-plane displacements for layers 1 to 4, respectively.}
\label{structure_AtABC}
\end{figure*}

\begin{figure*}
\centering
\begin{minipage}[b]{0.45\textwidth}
\centering
\includegraphics[width=\textwidth]{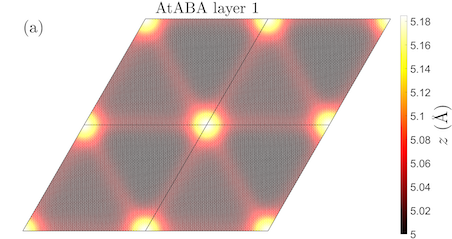}
\end{minipage}
\begin{minipage}[b]{0.45\textwidth}  
\centering 
\includegraphics[width=\textwidth]{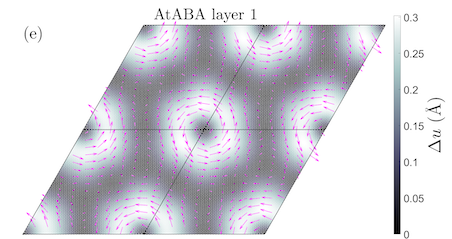}
\end{minipage}
\hfill
\vskip\baselineskip
\begin{minipage}[b]{0.45\textwidth}  
\centering 
\includegraphics[width=\textwidth]{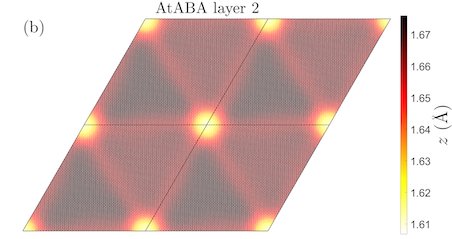}
\end{minipage}
\begin{minipage}[b]{0.45\textwidth}   
\centering 
\includegraphics[width=\textwidth]{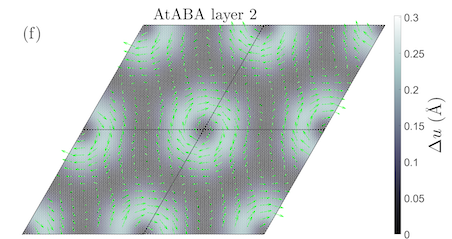}
\end{minipage}
\vskip\baselineskip
\begin{minipage}[b]{0.45\textwidth}
\centering
\includegraphics[width=\textwidth]{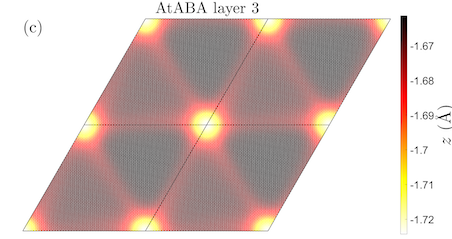}
\end{minipage}
\begin{minipage}[b]{0.45\textwidth}  
\centering 
\includegraphics[width=\textwidth]{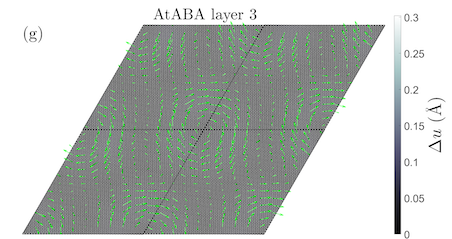}
\end{minipage}
\hfill
\vskip\baselineskip
\begin{minipage}[b]{0.45\textwidth}  
\centering 
\includegraphics[width=\textwidth]{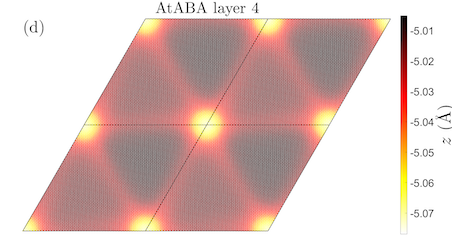}
\end{minipage}
\begin{minipage}[b]{0.45\textwidth}   
\centering 
\includegraphics[width=\textwidth]{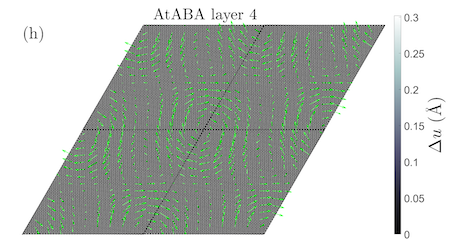}
\end{minipage}
\caption{Out-of-plane and in-plane relaxations of AtABA for a twist angle of $\theta=0.73\degree$. [(a)-(d)] Out-of-plane displacements for layers 1 to 4, respectively; [(e)-(h)] In-plane displacements for layers 1 to 4, respectively.}
\label{structure_AtABA}
\end{figure*}

\section{Acknowledgments}

ZG was supported through a studentship in the Centre for Doctoral Training on Theory and Simulation of Materials at Imperial College London funded by the EPSRC (EP/L015579/1). We acknowledge funding from EPSRC grant EP/S025324/1 and the Thomas Young Centre under grant number TYC-101. We acknowledge the Imperial College London Research Computing Service (DOI:10.14469/hpc/2232) for the computational resources used in carrying out this work. The Deutsche Forschungsgemeinschaft (DFG, German Research Foundation) is acknowledged for support through RTG 1995, within the Priority Program SPP 2244 “2DMP” and under Germany’s Excellence Strategy-Cluster of Excellence Matter and Light for Quantum Computing (ML4Q) EXC2004/1 - 390534769. We acknowledge support from the Max Planck-New York City Center for Non-Equilibrium Quantum Phenomena. Spin susceptibility calculations were performed with computing resources granted by RWTH Aachen University under projects rwth0496 and rwth0589.

\appendix

\begin{figure*}
\begin{minipage}[b]{0.3\textwidth}  
\centering 
\includegraphics[width=\textwidth]{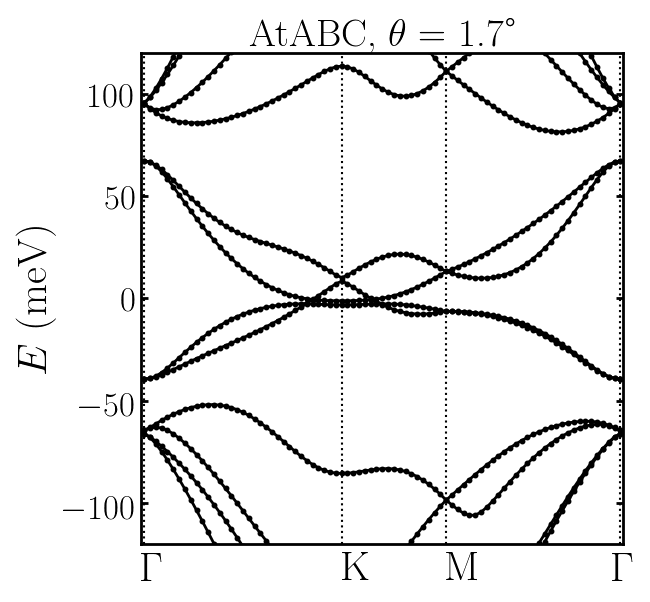}
\end{minipage}
\begin{minipage}[b]{0.3\textwidth}  
\centering 
\includegraphics[width=\textwidth]{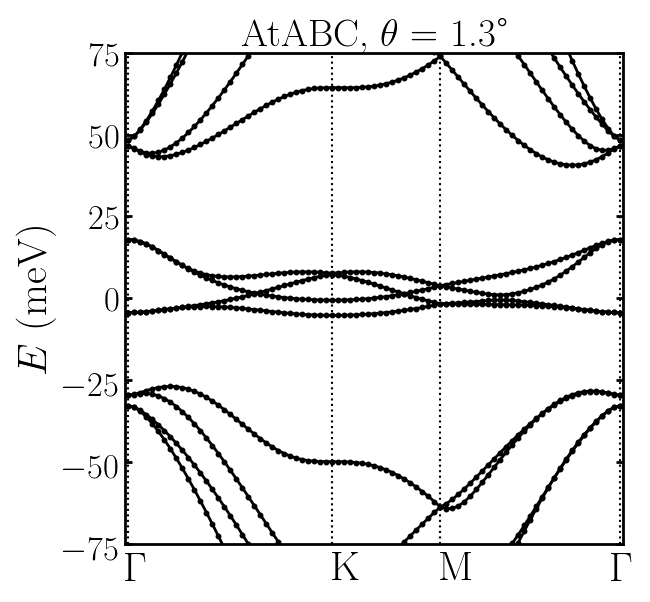}
\end{minipage}
\begin{minipage}[b]{0.3\textwidth}   
\centering 
\includegraphics[width=\textwidth]{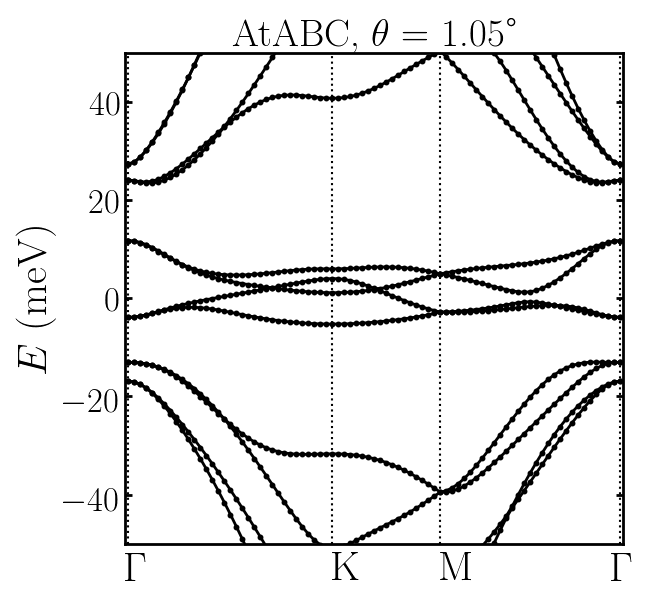}
\end{minipage}
\centering
\begin{minipage}[b]{0.3\textwidth}  
\centering 
\includegraphics[width=\textwidth]{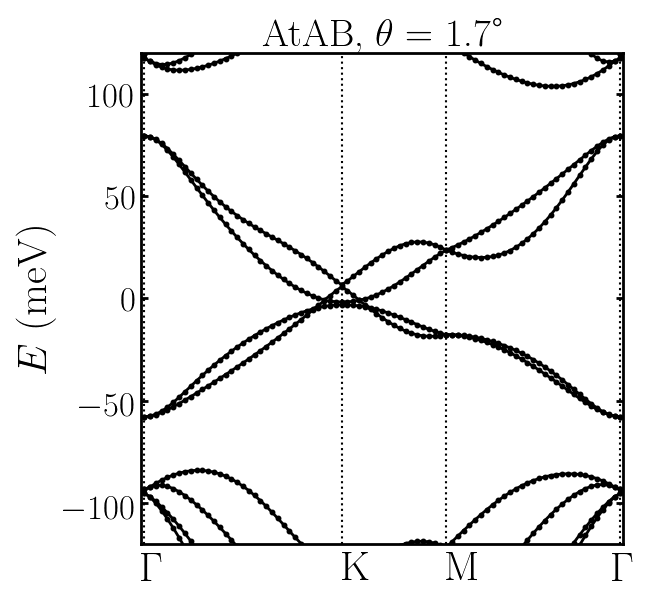}
\end{minipage}
\begin{minipage}[b]{0.3\textwidth}  
\centering 
\includegraphics[width=\textwidth]{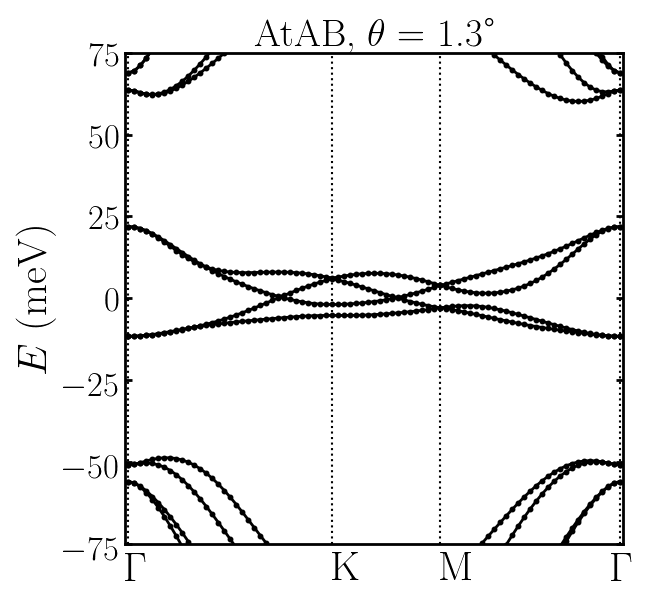}
\end{minipage}
\begin{minipage}[b]{0.3\textwidth}   
\centering 
\includegraphics[width=\textwidth]{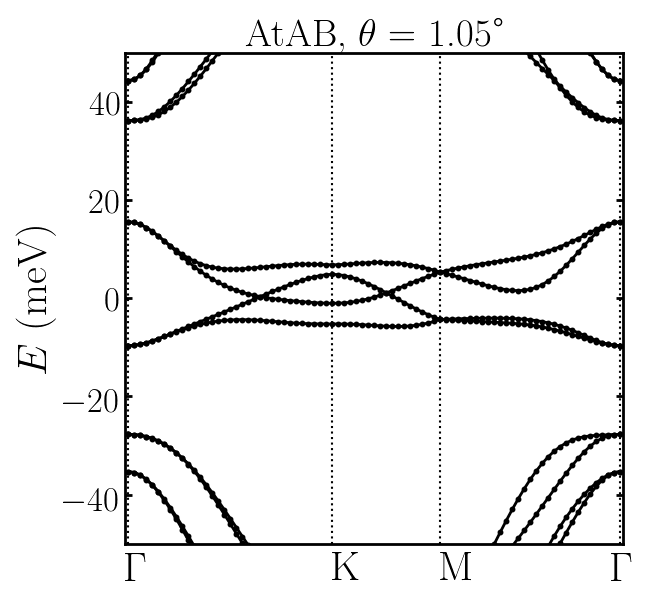}
\end{minipage}
\centering
\begin{minipage}[b]{0.3\textwidth}  
\centering 
\includegraphics[width=\textwidth]{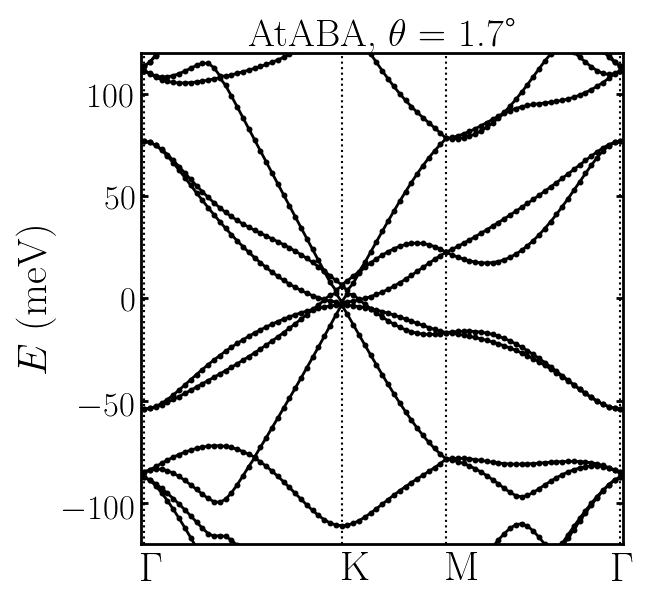}
\end{minipage}
\begin{minipage}[b]{0.3\textwidth}  
\centering 
\includegraphics[width=\textwidth]{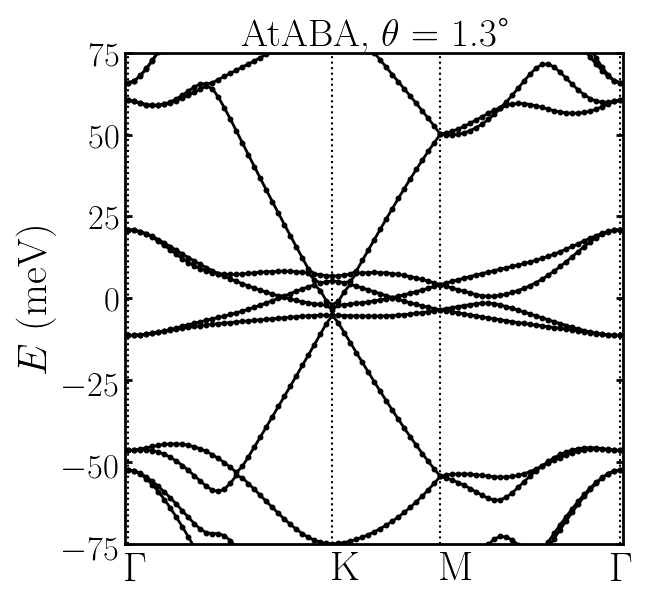}
\end{minipage}
\begin{minipage}[b]{0.3\textwidth}   
\centering 
\includegraphics[width=\textwidth]{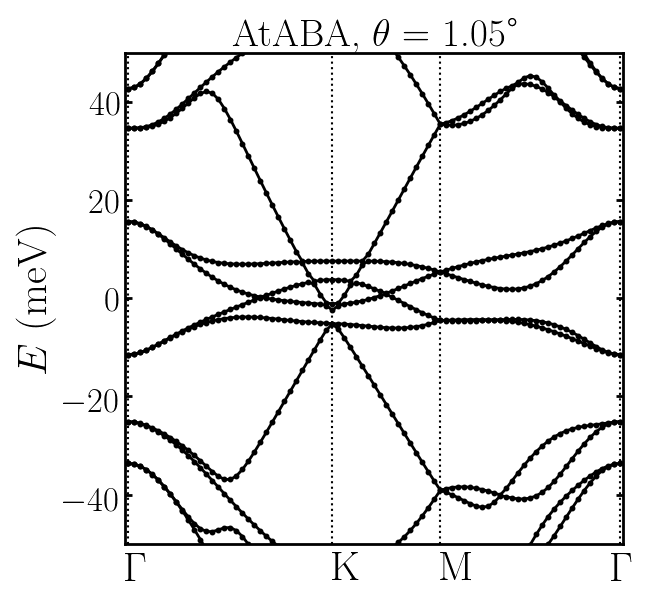}
\end{minipage}
\centering
\begin{minipage}[b]{0.3\textwidth}  
\centering 
\includegraphics[width=\textwidth]{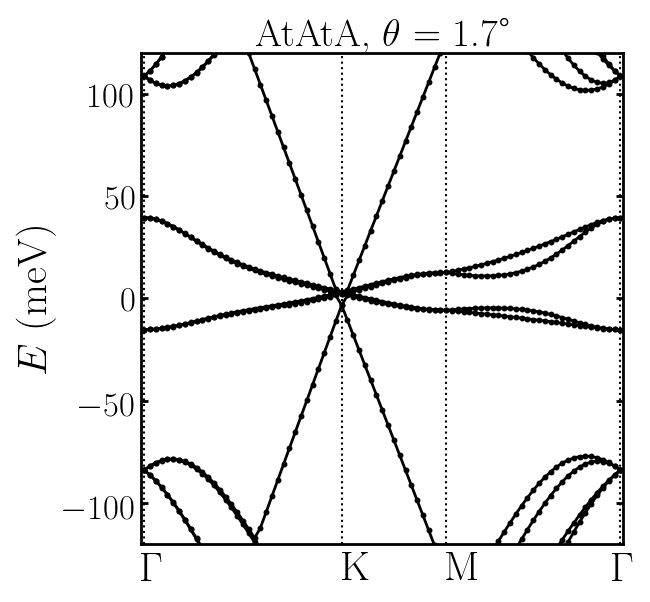}
\end{minipage}
\begin{minipage}[b]{0.3\textwidth}  
\centering 
\includegraphics[width=\textwidth]{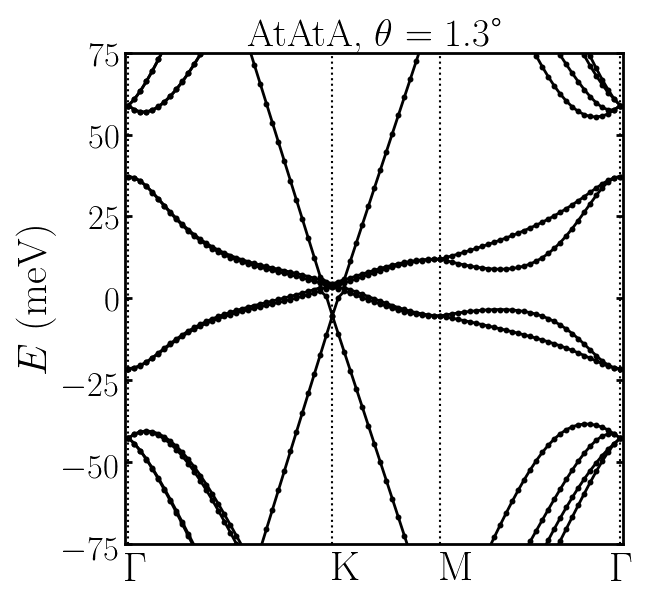}
\end{minipage}
\begin{minipage}[b]{0.3\textwidth}   
\centering 
\includegraphics[width=\textwidth]{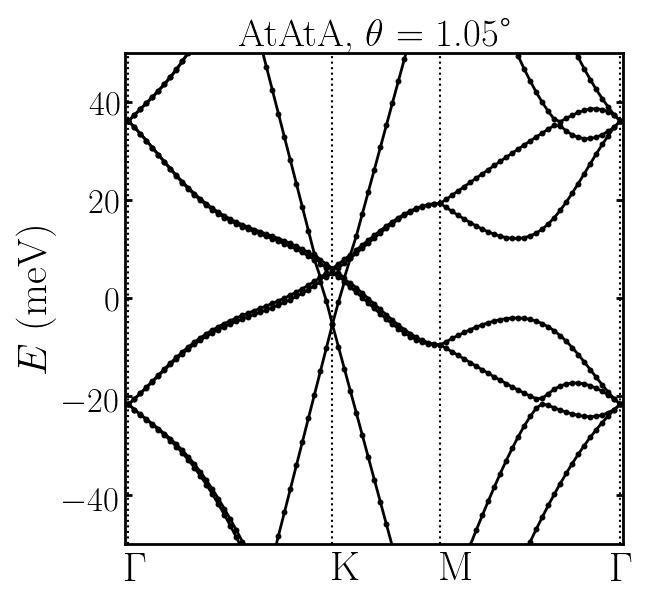}
\end{minipage}
\caption{Band structures from the atomistic tight-binding model along the high symmetry path. Relaxed atomic positions were used in each case.}
\label{BS_MORE}
\end{figure*}

\section{Appendix A: Atomic structure of mono-trilayer systems}

We study commensurate moir\'e unit cells of monolayer graphene twisted on trilayer graphene comprising of ABA or ABC stacking. The monolayer and trilayer are initially stacked directly on top of each other, and the top monolayer is rotated anticlockwise about an axis normal to the layers that passes through a carbon atom in the monolayer and the top layer of the trilayer. The moir\'e lattice vectors are $\textbf{R}_{1} = n\textbf{a}_{1} + m \textbf{a}_{2}$ and $\textbf{R}_{2} = -m\textbf{a}_{1} + (n + m) \textbf{a}_{2}$~\cite{LDE}, where $n$ and $m$ are integers that specify the moir\'e unit cell in terms of the graphene lattice vectors $\textbf{a}_{1}$ and $\textbf{a}_{2}$ with the lattice constant of graphene being $a_{0} = 2.42~\textrm{\AA}$.

In Figs.~\ref{structure_AtABC} and~\ref{structure_AtABA}, we display the relaxed structures of the studied mono-trilayer systems. We find that the relaxations of the graphene on trilayer graphene systems have some resemblance to twisted bilayer graphene (tBLG)~\cite{LREBM,STBBG,SETLA,CMLD,KDP} and also twisted double bilayer graphene (tDBLG)~\cite{Xia2020tDBLG,haddadi2019moir}.

Both AtABC and AtABA exhibit similar lattice reconstruction, as shown in Figs.~\ref{structure_AtABC} and~\ref{structure_AtABA}, with the relaxation features being analogous to tBLG and tDBLG~\cite{Xia2020tDBLG}. In both of these structures, the twisted graphene layer (layer 1) and the graphene layer that is in contact with the twisted layer (layer 2) undergo the most significant relaxations. These two layers form a ``tBLG unit", and the relaxation effects of these layers in AtABC and AtABA can be seen to be analogous. Namely, there are peaks in the $z$-displacement in the AA regions of these layers, owing to the unfavorable stacking order; the in-plane displacements have an opposite sense in each layer, such that the AB stacking order is increased relative to AA.

In layer 3 of both structures, the magnitudes of the $z$-displacements are similar to those of layer 2. However, the in-plane displacements on layer 3 are significantly less pronounced than on layers 1/2. This indicates that the ABC and ABA stacking is not perfectly retained throughout the whole moir\'e unit cell. In layer 4 the in-plane and out-of-plane relaxations are similar to those of layer 3, but the magnitudes of the displacements are again even smaller.

\section{Appendix B: Electronic structure from tight-binding}

The electronic structure was investigated with an atomistic tight-binding model, which is a reliable method for determining the electronic structure of graphene-based moir\'e materials. In the atomistic tight-binding formalism, the Hamiltonian is given by
\begin{equation}
\mathcal{\hat{H}} = \sum_{i}\varepsilon_{i}\hat{c}^{\dagger}_{i}\hat{c}_{i} + \sum_{ij}[t(\textbf{t}_{i} - \textbf{t}_{j})\hat{c}^{\dagger}_{j}\hat{c}_{i} + \text{H.c.}].
\label{eq:H}
\end{equation}
Here $\hat{c}^{\dagger}_{i}$ and $\hat{c}_{i}$ are, respectively, the electron creation and annihilation operators associated with the p$_{z}$-orbital on atom $i$. The $\varepsilon_{i}$ is the on-site energy of the p$_{z}$-orbitals, which is used to fix the Fermi energy at 0~eV (and later to include Hartree interactions). The hopping parameters $t(\textbf{t}_{i} - \textbf{t}_{j})$ between atoms $i$ and $j$ (located at $\mathbf{t}_{i/j}$) are determined using the Slater-Koster rules
\begin{equation}
t(\textbf{r}) = \gamma_1e^{-(|\textbf{r}| - d)/\delta_0}\cos^{2}\varphi + \gamma_0e^{-(|\textbf{r}| - a)/\delta_0}\sin^{2}\varphi.
\end{equation}
Here $\gamma_{1} = 0.48$ eV and $\gamma_{0} = -2.70$ eV correspond to $\sigma$ and $\pi$ hopping between p$_{z}$-orbitals, respectively. The carbon-carbon bond length is $a = 1.397~\textrm{\AA}$ and the interlayer separation parameter is taken to be $d = 1.36a_0~\textrm{\AA}$. The decay parameter of the hoppings is set to $\delta_0 = 0.184a_0$. The angle-dependence of hoppings are captured through $\varphi$, which is the angle corresponding between the $z$-axis and the vector connecting atoms $i$ and $j$. Hoppings between carbon atoms whose distance is larger than the cutoff $10~\textrm{\AA}$ are neglected. 

In Fig.~\ref{BS_MORE} we display additional band structures for AtABC, AtABA, AtAB and AtAtA as a function of twist angle. There are similarities between the AtAB and AtABC band structures, as discussed in the main text. An analogy can also be drawn between AtABA and AtAtA, from the fact that both systems have a Dirac cone intersecting the flat moir\'e bands.

\section{Appendix C: Density of states}

\begin{figure}[ht]
\centering
\begin{minipage}[b]{0.4\textwidth}  
\centering 
\includegraphics[width=\textwidth]{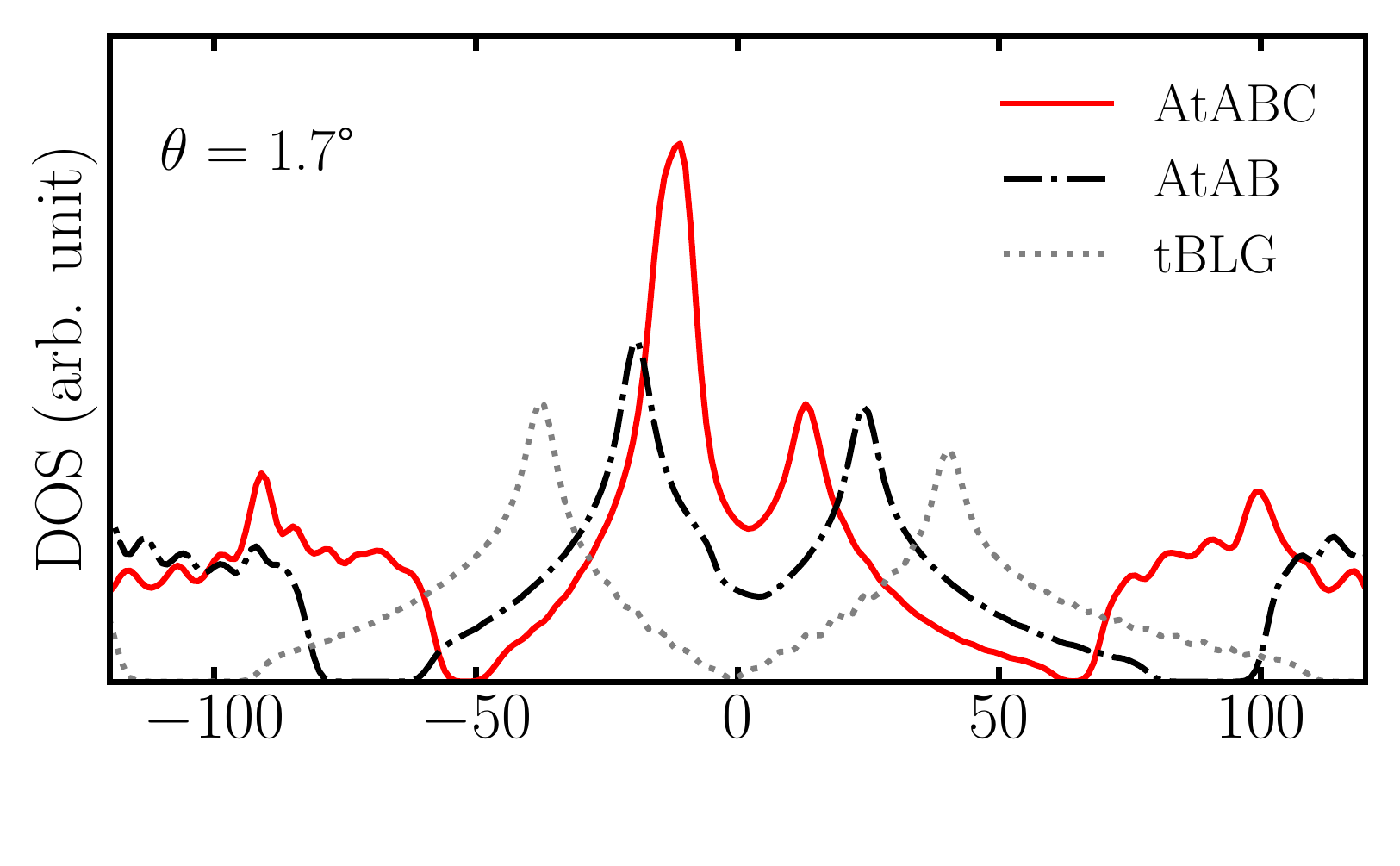}
\end{minipage}
\begin{minipage}[b]{0.4\textwidth}  
\centering 
\includegraphics[width=\textwidth]{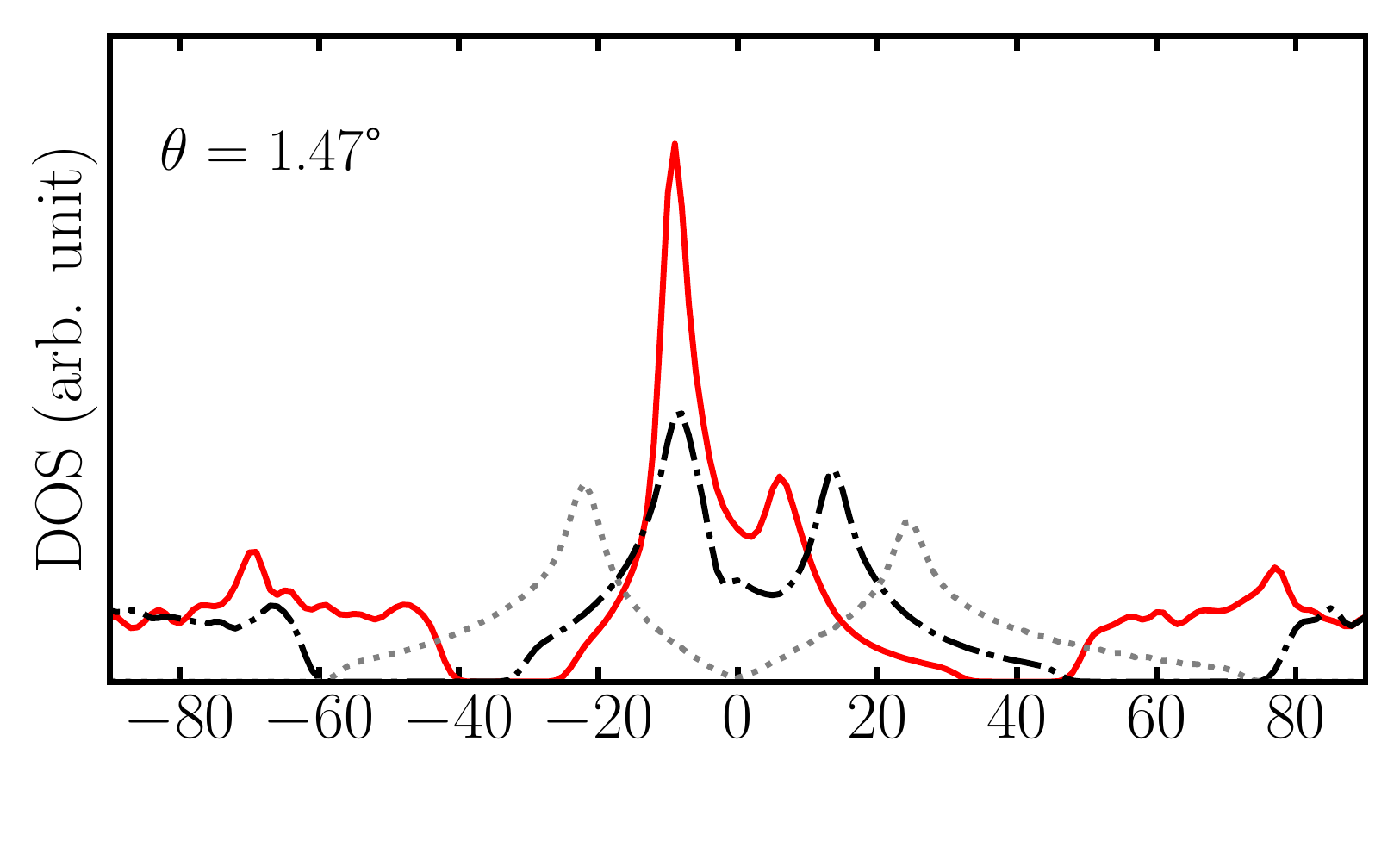}
\end{minipage}
\begin{minipage}[b]{0.4\textwidth}  
\centering 
\includegraphics[width=\textwidth]{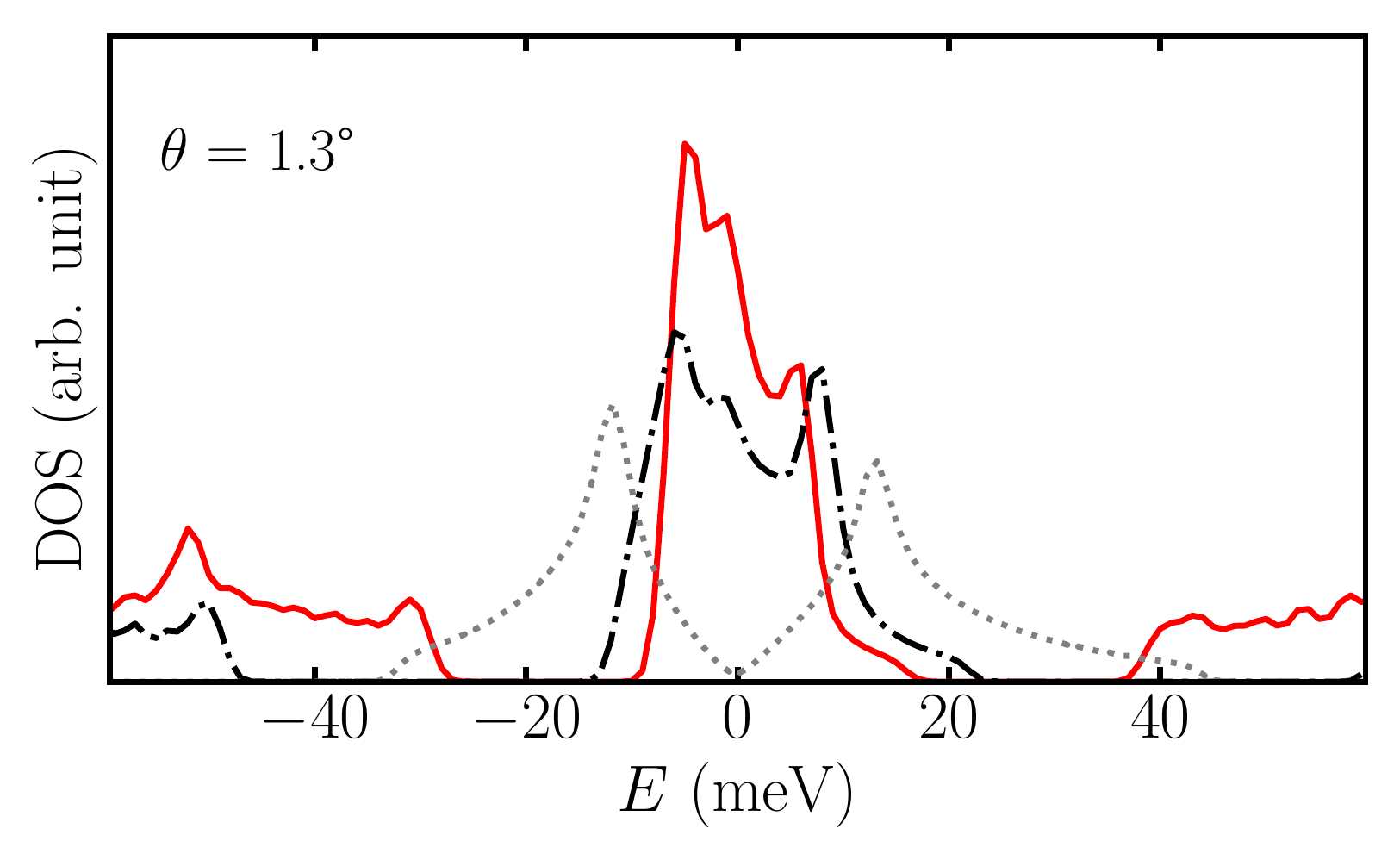}
\end{minipage}
\caption{Density of states (DOS) as a function of energy for AtABC, AtAB and tBLG at a twist angle of 1.7$\degree$, 1.47$\degree$ and 1.3$\degree$. The zero of  energy  is  set  to  the  Fermi  level  at  charge  neutrality  for each system, and the zero in the DOS is at the bottom of the $y$-axis.}
\label{DOS_Appendix}
\end{figure}

To calculate the density of states, we employ a Gaussian broadening scheme with a 31$\times$31 k-point Monkhorst-Pack grid which includes the $\Gamma$-point. A broadening parameter of 32, 18, 14, 10~meV were used for twist angles 2.0$\degree$, 1.7$\degree$, 1.47$\degree$ and 1.3$\degree$, respectively. In Fig.~\ref{DOS_Appendix} the latter twist angles are shown.

\begin{figure*}
\centering
\begin{minipage}[b]{0.32\textwidth}  
\centering 
\includegraphics[width=\textwidth]{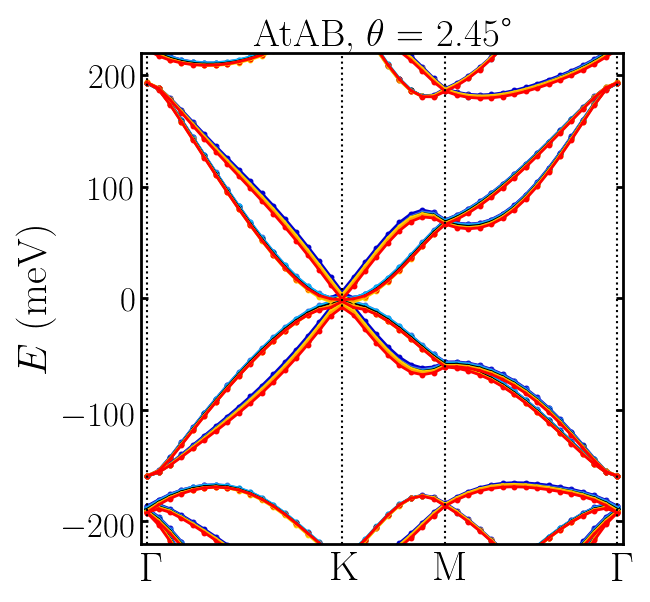}
\end{minipage}
\begin{minipage}[b]{0.32\textwidth}  
\centering 
\includegraphics[width=\textwidth]{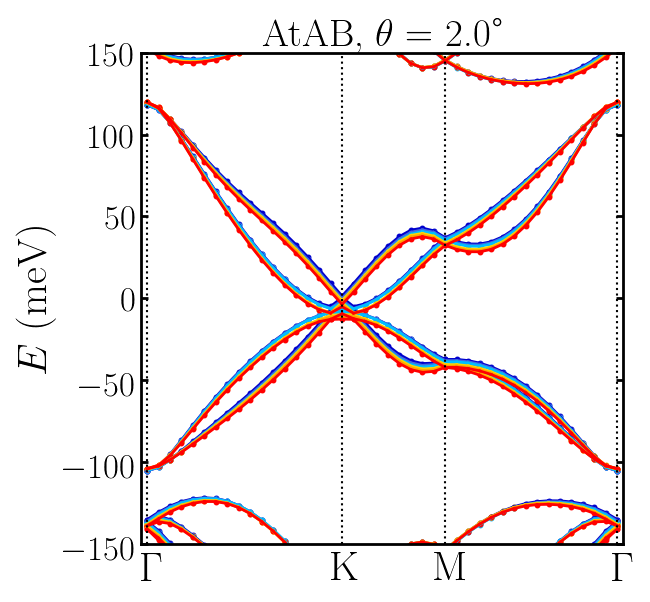}
\end{minipage}
\begin{minipage}[b]{0.32\textwidth}  
\centering 
\includegraphics[width=\textwidth]{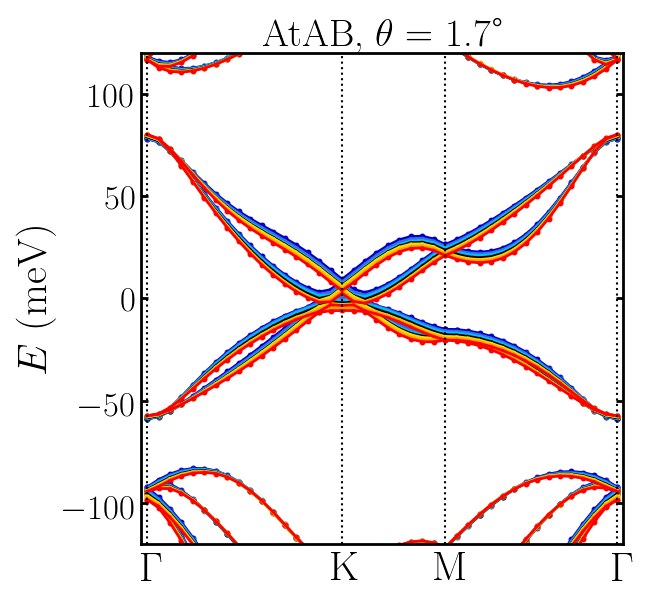}
\end{minipage}
\centering
\begin{minipage}[b]{0.32\textwidth}  
\centering 
\includegraphics[width=\textwidth]{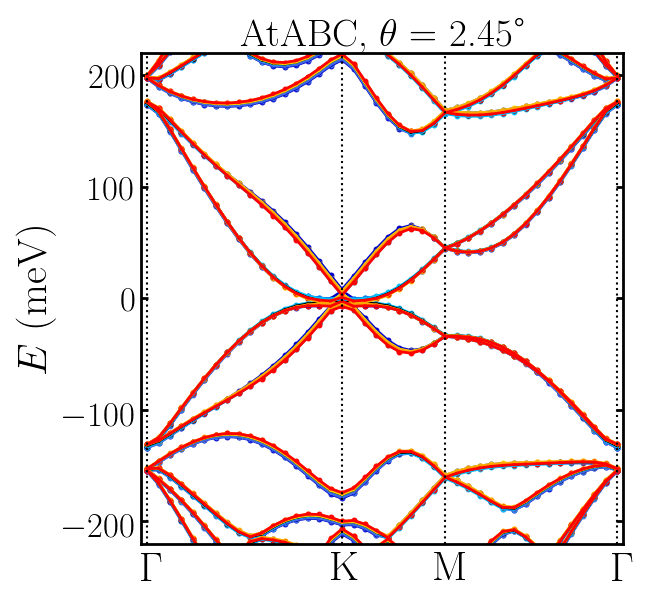}
\end{minipage}
\begin{minipage}[b]{0.32\textwidth}  
\centering 
\includegraphics[width=\textwidth]{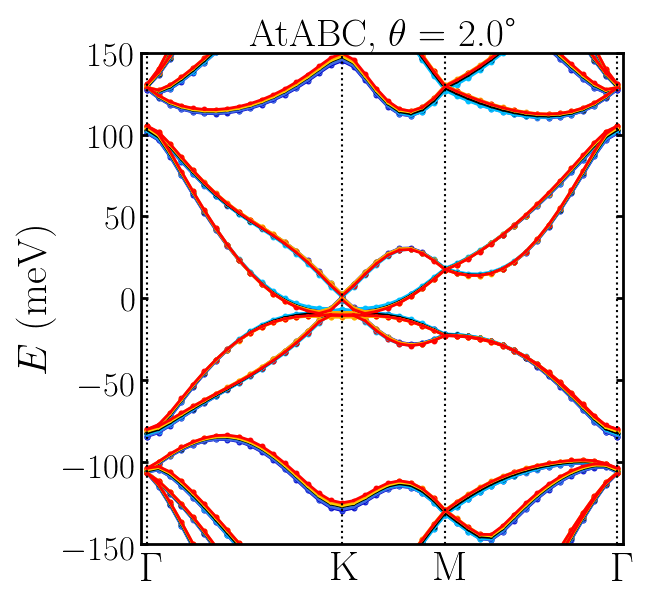}
\end{minipage}
\begin{minipage}[b]{0.32\textwidth}  
\centering 
\includegraphics[width=\textwidth]{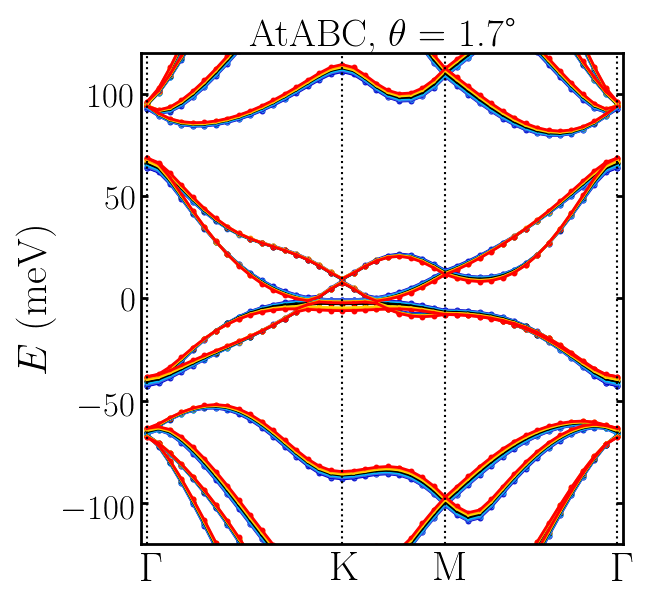}
\end{minipage}
\caption{Hartree theory band structures of AtAB and AtABC at 2.45$\degree$, 2.0$\degree$ and 1.7$\degree$ for integer doping levels per moir\'e unit cell from $\nu = 3$ (blue) to $\nu = -3$ (red).}
\label{BS_Hart}
\end{figure*}

\begin{figure*}[ht]
\centering
\begin{minipage}[b]{0.49\textwidth}
    \makebox[8pt]{\footnotesize\bfseries(a)}
    \includegraphics[valign=t,width=0.9\textwidth]{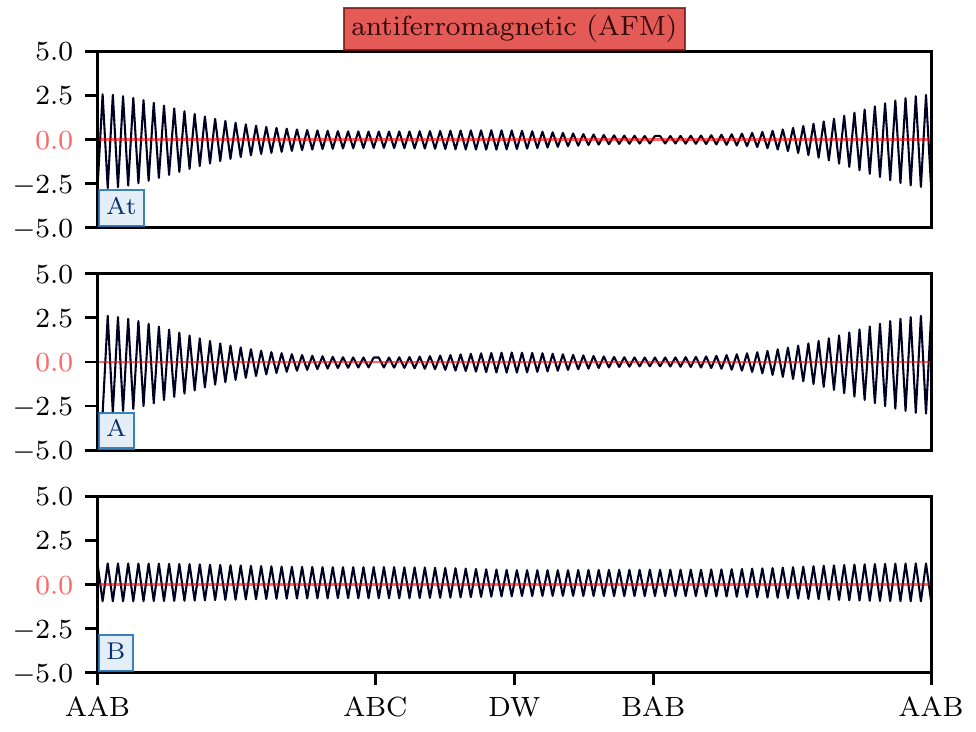}
\end{minipage}
\begin{minipage}[b]{0.49\textwidth}
    \makebox[8pt]{\footnotesize\bfseries(b)}
    \includegraphics[valign=t,width=0.9\textwidth]{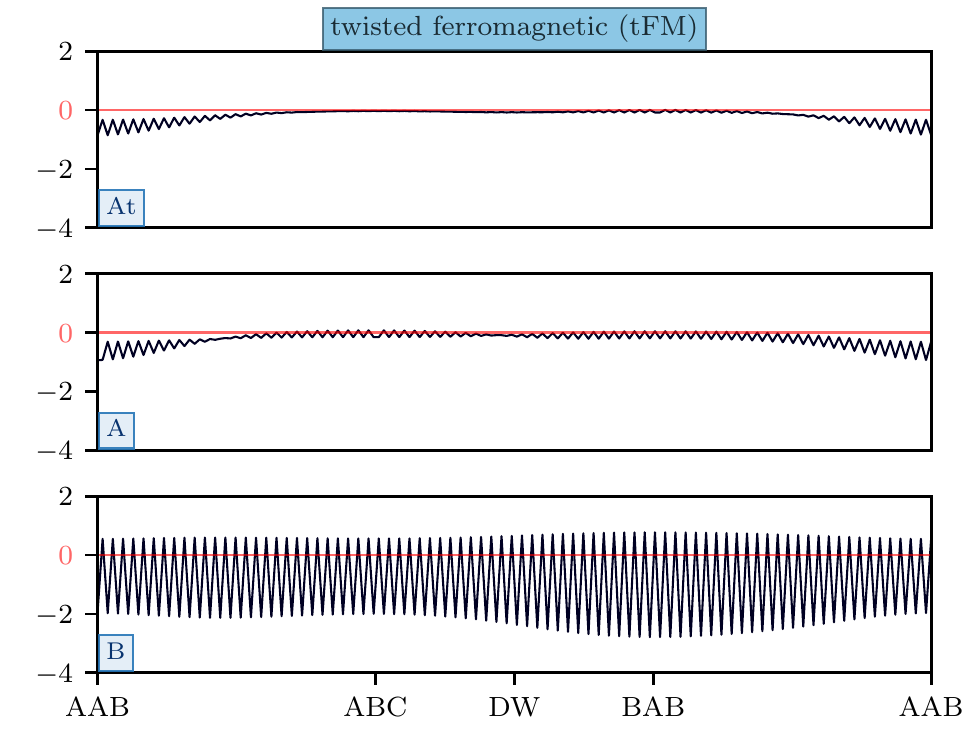}
\end{minipage} \\[5pt]
\begin{minipage}[b]{0.49\textwidth}
    \makebox[8pt]{\footnotesize\bfseries(c)}
    \includegraphics[valign=t,width=0.9\textwidth]{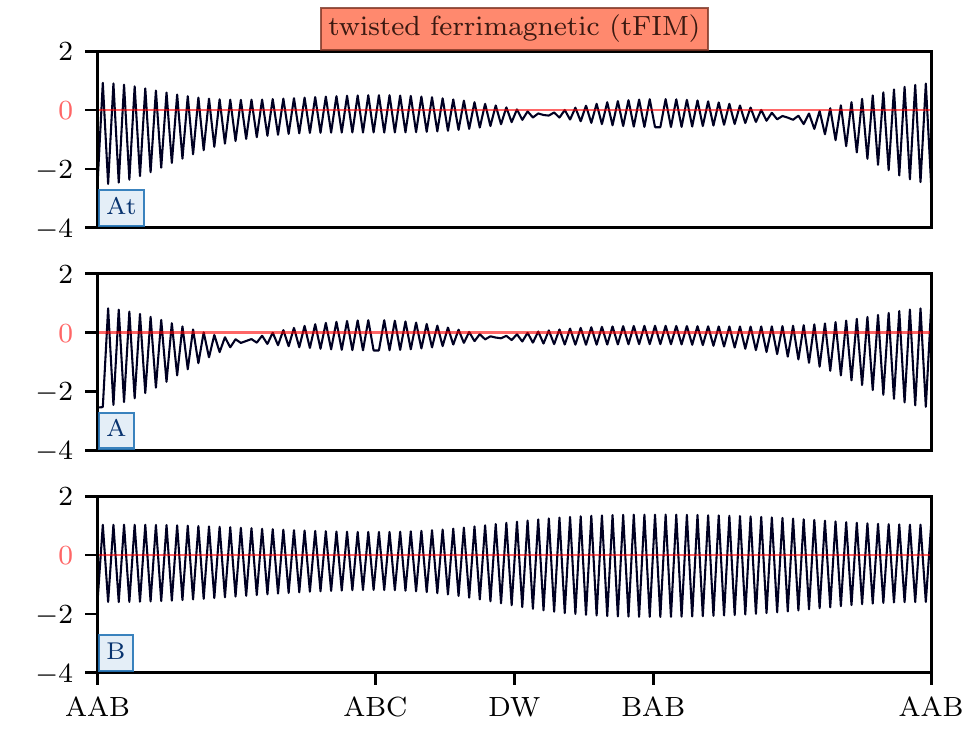}
\end{minipage}
\begin{minipage}[b]{0.49\textwidth}
    \makebox[8pt]{\footnotesize\bfseries(d)}
    \includegraphics[valign=t,width=0.9\textwidth]{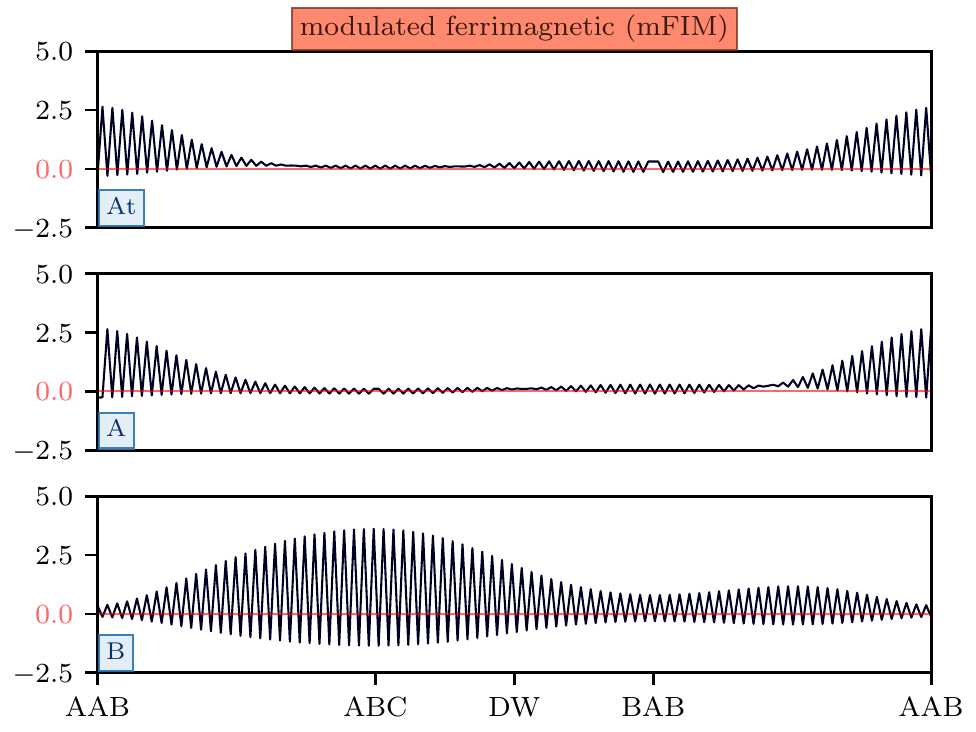}
\end{minipage}
\caption{Linecut of different magnetic ordering ($y$-axis is the normalized eigenvector corresponding to eigenvalue $\chi^0$ at the given twist angle and doping level) through the long diagonal of the moir\'e unit cell for the three layers of AtAB graphene at various angles and fillings. The local stacking sequence is shown at the bottom of each panel, where DW stands for the domain wall region of the moir\'e pattern.
{\bfseries{}(a)}: $\theta = 1.20\degree, \nu = 2$ -- ordering tendency that is strongly anti-ferromagnetic in each layer with more pronounced localization in the twisted two layers.
{\bfseries{}(b)}: $\theta = 1.20\degree, \nu = -3$ -- ordering tendency that is mainly ferromagnetic in the twisted layers with reminiscent anit-ferromagnetic order, with ferrimagnetic order in the lower layer.
{\bfseries{}(c)}: $\theta = 1.25\degree, \nu = 3$ -- ordering where there is modulated ferrimagnetic order in the layers with a relative twist angle, and uniform anti-ferromagnetic order in the lower untwisted layer.
{\bfseries{}(d)}: $\theta = 1.16\degree, \nu = 1$ -- order that has strongly modulated ferrimagnetic order in all layers.}
\label{fig:linecut-ab}
\end{figure*}

\begin{figure*}
\begin{minipage}[b]{0.49\textwidth}
\centering 
    \makebox[8pt]{\footnotesize\bfseries(a)}
    \includegraphics[valign=t,scale=0.9]{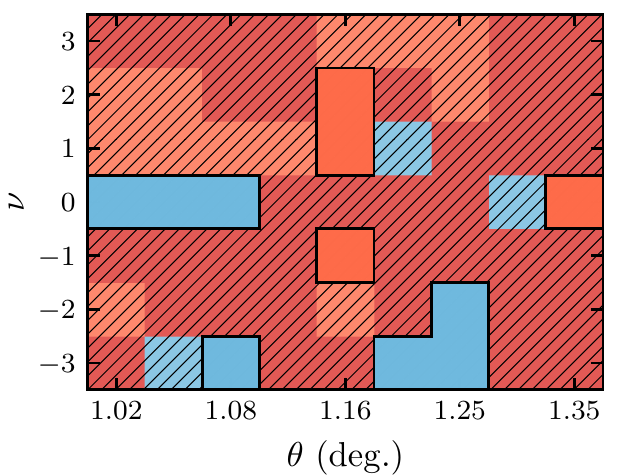}
\end{minipage}
\begin{minipage}[b]{0.49\textwidth}
\centering 
    \makebox[8pt]{\footnotesize\bfseries(b)}
    \includegraphics[valign=t,scale=0.9]{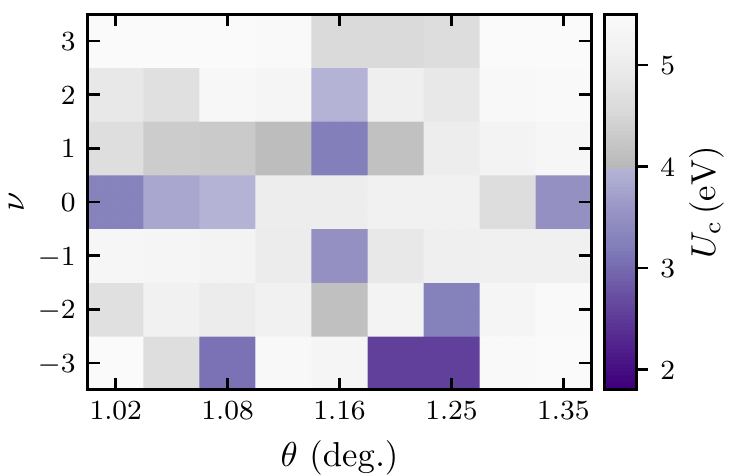}
\end{minipage}
\caption{Magnetic phase diagram of AtAB graphene for twist angles ($\theta$) around the first magic angle and integer fillings ($\nu$).
{\bfseries{}(a)}: Types of magnetic order: Blue corresponds to ferromagnetic order, orange corresponds to ferrimagnetic order, and red corresponds to anti-ferromagnetic order, examples of which can be seen in Fig.~\ref{fig:linecut-ab}. We have hatched out the where the values of $\Ucrit$ are lower than 4~eV.
{\bfseries{}(b)}: Critical interaction strength $\Ucrit$ required for the onset of magnetic instabilities.}
\label{fig:atab}
\end{figure*}

\section{Appendix D: Hartree interactions}

The long-ranged electron-electron interaction contribution to the Hamiltonian can be included through 
\begin{equation}
\varepsilon_i = \int d\textbf{r} \phi_z^2(\textbf{r}-\boldsymbol{\tau}_i)\VHr,
\end{equation}
where $\phi_z(\textbf{r})$ is the p$_z$ orbital of the carbon atoms and \VHr\ is the Hartree potential. The Hartree potential is determined from the electron density $n(\mathbf{r})$ and the screened electron-electron interaction $W(\mathbf{r})$, as seen by
\begin{equation}
\VHr = \int d\textbf{r}' W(\textbf{r}-\textbf{r}') [n(\textbf{r}') - n_0(\textbf{r}')],
\label{eq:VH}
\end{equation} 
where $n_0(\mathbf{r})$ is a reference electron density of the uniform system. The electron density is determined through
\begin{equation}
n(\textbf{r}) = \sum_{n\textbf{k}} f_{n\textbf{k}} |\psi_{n\textbf{k}}(\textbf{r})|^2
\label{eq:el_den}
\end{equation}
where $\psi_{n\textbf{k}}(\textbf{r})$ is the Bloch eigenstate of the atomistic tight-binding model, with subscripts $n$ and $\mathbf{k}$ denoting the band index and crystal momentum, respectively; $N_{\textrm{k}}$ is the number of k-points in the summation of the electron density, and $f_{n\textbf{k}}=2\Theta(\varepsilon_\mathrm{F}-\varepsilon_{n\mathbf{k}})$ is the spin-degenerate occupancy of state $\psi_{n\mathbf{k}}$ with eigenvalue $\varepsilon_{n\mathbf{k}}$ (where $\varepsilon_\mathrm{F}$ is the Fermi energy). Inserting the Bloch states in Eq.~\eqref{eq:el_den} gives
\begin{equation}
n(\textbf{r}) =\sum_{j}n_j\chi_j(\textbf{r}),
\end{equation}
where $\chi_j(\textbf{r}) = \sum_\mathbf{R} \phi_{z}^2(\textbf{r}-\boldsymbol{\tau}_j-\textbf{R})$ (with $\mathbf{R}$ denoting the moir\'e lattice vectors) and the total number of electrons on the $j$-th p$_z$-orbital in the unit cell being determined by $n_j = \sum_{n\mathbf{k}} f_{n\textbf{k}}|c_{n\mathbf{k}j}|^2/N_{\textrm{k}}$, with $c_{n\mathbf{k}j}$ denoting the coefficients of the eigenvectors of the tight-binding model. 

The reference density is taken to be that of a uniform system, $n_0(\textbf{r}) = \bar{n} \sum_j \chi_j(\textbf{r})$, where $\bar{n}$ is the average of $n_j$ over all atoms in the unit cell, which is related to the filling per moir\'e unit cell $\nu$ through $\bar{n}=1+\nu/N$, where $N$ is the total number of atoms in a moir\'e unit cell~\cite{Rademaker2019}. This reference density is taken to prevent overcounting the intrinsic graphene Hartree contribution which should be included in the hopping parameters of Eq.~\eqref{eq:H}.

In experiments, there is often a metallic gates above and below the moir\'e material, with a hexagonal boron nitride (hBN) substrate separating the gates from moir\'e materials. These metallic gates add/remove electrons from moir\'e material and can also create electric fields across the system. These gates also screen the electron interactions in moir\'e material, and taking this effect into account has been shown to be important in tBLG~\cite{PHD_3}. Therefore, we utilize a double metallic gate screened interaction
\begin{equation}
W(\textbf{r}) = \dfrac{e^2}{4\pi\epsilon_0\epsilon_\mathrm{bg}}\sum_{m=-\infty}^{\infty} \dfrac{(-1)^{m}}{\sqrt{|\textbf{r}|^2 + (2m\xi)^2}},
\label{WMG}
\end{equation} 
where $\xi$ is the thickness of the hBN dielectric substrate, with dielectric constant $\epsilon_\mathrm{bg}$, separating tDBLG from the metallic gate on each side~\cite{MGS,PHD_1,PHD_3}. We set $\xi = 10$~nm  and the value of $\epsilon_\mathrm{bg} = 4$ for all calculations. 

In our atomistic model, we neglect contributions to the electron density from overlapping p$_{z}$-orbitals that do not belong to the same carbon atom, which is equivalent to treating $\phi^2_z(\textbf{r})$ as a delta-function. Therefore, we calculate the Hartree on-site energies using
\begin{equation}
\varepsilon_i = \sum_{j\textbf{R}}(n_j - \bar{n}) W_{\textbf{R}ij},
\label{eq:H_pot}
\end{equation}
where $W_{\textbf{R}ij}=W(\textbf{R}+\boldsymbol{\tau}_j-\boldsymbol{\tau}_i)$. If $\textbf{R}=0$ and $i=j$, we set $W_{0,ii}=U/\epsilon_\mathrm{bg}$ with $U=17$~eV~\cite{SECI}. 

To obtain a self-consistent solution of the equations, we use a $8\times 8$ k-point grid to sample the first Brillouin zone to converge the density in Eq.~\eqref{eq:el_den} and we sum over an $11\times 11$ supercell of moir\'e unit cells to converge the on-site energy of Eq.~\eqref{eq:H_pot}. Linear mixing of the electron density is performed with a mixing parameter of 0.1 or less (i.e., the addition of 10 percent of the new potential to 90 percent of the potential from the previous iteration). Typically, the Hartree potential converges to an accuracy of better than 0.1~meV per atom within 100 iterations. For doping levels where the moir\'e material is metallic, smaller mixing values and more iterations are sometimes needed to reach this convergence threshold as the lack of smearing causes states very close in energy to flick between being occupied and un-occupied in consecutive iterations.

In Fig.~\ref{BS_Hart} we show the quasiparticle band structure for AtAB and AtABC for a twist angles of 2.45$\degree$, 2.0$\degree$ and 1.7$\degree$ at integer doping levels from $\nu = 3$ to $\nu = -3$. Overall, we find that the electronic structure is rather insensitive to the long-ranged electron-electron interactions. The scale of the Hartree potential is quite small (5-10~meV), and there is not a clean localization of states which can give rise to significant band distortions. The changes in the band structure predominantly come from layer-dependent differences, rather than in-plane variations. Therefore, we can safely neglect the long-ranged Hartree interactions and just use the tight-binding model for magnetic calculations.

\section{Appendix E: Random phase approximation}

Following Refs.~\onlinecite{LK_CH,PHD_6} to analyze the magnetic ordering tendencies of graphitic moir\'e systems, we calculate the spin susceptibility $\chi_{ij}(\bvec q,q_0)$ in its long-wavelength, static limit $\bvec q = q_0 = 0$:
\begin{equation}
    \hat\chi=\hat\chi(\bvec q=0,q_0=0)=\frac{T}{N_{\bvec{k}}} \sum_{\bvec{k},k_0} \hat G(\bvec k,k_0) \hadamard \hat G^T(\bvec k,k_0).
\end{equation}
The Matsubara Green's function reads $\hat G(\bvec k,k_0) = (ik_0 - \hat H(\bvec k))^{-1}$ with the non-interacting part of the Hamiltonian $\hat H(\bvec k)$.
Since we approximate the interacting part of the Hamiltonian by a local Hubbard interaction, the renormalized interaction reads
\begin{equation}
    \hat W = \frac{U^2\hat\chi}{\mathds{1}+U\hat\chi}.
\end{equation}
Employing Stoner's criterion, we find an ordered state if the smallest eigenvalue $\chi^0$ of the matrix $\hat\chi$ reaches $-1/U$, or, vice versa, we can investigate the critical interaction strength $\Ucrit = -1/\chi^0$ below which the system will go into an ordered state. The eigenvector corresponding to the eigenvalue $\chi^0$ is proportional to the system's magnetization in its ordered state. The numerical evaluation procedure is identical to the one presented in Ref.~\onlinecite{PHD_6} -- we use $N_{k_0} = 500$ Matsubara frequencies and $N_{\bvec k}=24$ momentum points at a temperature of $T=10^{-4}\,\mathrm{eV}$.

Similarly to AtABC, the leading magnetic instabilities found in AtAB graphene can be classified as modulated AFM order, modulated FM in the top two layers and FIM in the bottom layer, and two types of FIM order. The line-cuts of the orderings for magic-angle AtAB can be found in Fig.~\ref{fig:linecut-ab}.

In Fig.~\ref{fig:atab} we shown the magnetic phase diagram of AtAB around the magic angle at integer doping levels in the flat bands. Similar to AtABC, we find that AtAB does not have as strong of a tendency to magnetic order as tBLG. Compared with AtABC graphene, the $\Ucrit$ is low mostly at $\nu=0$ and $\nu=-3$ [Fig.~\ref{fig:atab}~{\bfseries{}(b)}] with FM orders being more prominent in the magnetic phase diagram [Fig.~\ref{fig:atab}~{\bfseries{}(a)}].

\bibliographystyle{apsrev4-1}
\bibliography{REF}

\end{document}